\documentclass[final,3p,times,authoryear]{elsarticle}

\usepackage{amssymb,amsmath}
\usepackage{adjustbox,booktabs,multirow}
\usepackage{lineno}

\journal{ArXiv}

\usepackage{enumitem}
\usepackage{url}
\usepackage{graphicx}
\usepackage{tabularx}
\usepackage{xcolor, soul}
\usepackage[T1]{fontenc}
\usepackage{pdflscape}
\usepackage{pdfpages}
\usepackage{color}
\definecolor{RoyalBlue}{cmyk}{1, 0.50, 0, 0}

\begin{document}

\newcommand{\dtmref}{DTM\textsubscript{reference}}
\newcommand{\dsmref}{DSM\textsubscript{reference}}
\newcommand{\demref}{DEM\textsubscript{reference}}
\newcommand{\demtest}{DEM\textsubscript{test}}

\begin{frontmatter}

\indent \textcolor{RoyalBlue}{Cite as: Bielski, C.; L{\'o}pez-V{\'a}zquez, C.; Grohmann, C.H.; Guth. P.L.; Hawker, L.; Gesch, D.; Trevisani, S.; Herrera-Cruz, V.; Riazanoff, S.; Corseaux, A.; Reuter, H.; Strobl, P., 2024. Novel approach for ranking DEMs: Copernicus DEM improves one arc second open global topography. IEEE Transactions on Geoscience \& Remote Sensing. \url{https://doi.org/10.1109/TGRS.2024.3368015}}  \\

 \vspace{15pt} 



\title{Novel approach for ranking DEMs: Copernicus DEM improves one arc second open global topography}

\author[eox]{Conrad Bielski}
\ead{conrad.bielski@eoxplore.com}
\ead[url]{https://orcid.org/0000-0002-2804-5150}

\author[uou]{Carlos L{\'o}pez-V{\'a}zquez}
\ead{carloslopez@uni.ort.edu.uy}
\ead[url]{https://orcid.org/0000-0002-8444-1510}

\author[usp]{Carlos H. Grohmann\corref{cor1}}
\ead{guano@usp.br}
\ead[url]{https://orcid.org/0000-0001-5073-5572} 

\author[usna]{Peter L. Guth}
\ead{pguth@usna.edu}
\ead[url]{https://orcid.org/0000-0002-6150-542X}

\author[bristol]{Laurence Hawker}
\ead{laurence.hawker@bristol.ac.uk}
\ead[url]{https://orcid.org/0000-0002-8317-7084}

\author[usgs]{Dean Gesch}
\ead{gesch@usgs.gov}
\ead[url]{https://orcid.org/0000-0002-8992-4933}

\author[iuav]{Sebastiano Trevisani}
\ead{strevisani@iuav.it}
\ead[url]{https://orcid.org/0000-0001-8436-7798}

\author[airbus]{Virginia Herrera-Cruz}
\ead{virginia.herrera@airbus.com}
\ead[url]{https://orcid.org/0000-0003-1746-1471}

\author[visioterra]{Serge Riazanoff}
\ead{serge.riazanoff@visioterra.fr}
\ead[url]{https://orcid.org/0000-0001-6399-7299}

\author[visioterra]{Axel Corseaux}
\ead{axel.corseaux@visioterra.fr}
\ead[url]{https://orcid.org/0009-0009-4960-7472}

\author[eurostat]{Hannes I. Reuter}
\ead{hannes.reuter@ec.europa.eu}
\ead[url]{https://orcid.org/0000-0001-6336-7801}

\author[jrc]{Peter Strobl}
\ead{Peter.STROBL@ec.europa.eu}
\ead[url]{https://orcid.org/0000-0003-2733-1822}


\cortext[cor1]{Corresponding author}

\address[eox]{EOXPLORE, Gaugrafenstr. 8, Oberhaching, 8204, Bavaria, Germany}

\address[uou]{LatinGEO Lab IGM+ORT, Universidad ORT URUGUAY, Cuareim 1451, Montevideo, 11100, Montevideo, Uruguay}

\address[usp]{Universidade de S\~{a}o Paulo, Av. Prof. Luciano Gualberto, 1289 , S\~{a}o Paulo, 05508-010 , SP, Brazil}

\address[usna]{Department of Ocean and Atmospheric Sciences, US Naval Academy, Annapolis, 21402, MD, USA (retired)}

\address[bristol]{School of Geographical Sciences, University of Bristol, BS8 1SS Bristol, UK}

\address[usgs]{U.S. Geological Survey, Earth Resources Observation and Science Center, 57198 Sioux Falls, USA}

\address[iuav]{University IUAV of Venice, 30135 Venice, Italy}

\address[airbusUniversity IUAV of Venice, 30135 Venice, Italy]{Airbus Defence and Space, 88039 Friedrichshafen, Germany}

\address[visioterra]{VisioTerra, 77420 Champs-sur-Marne, France}

\address[eurostat]{European Commission, Eurostat, L-2721, Luxembourg}

\address[jrc]{European Commission, DG Joint Research Centre, 21027 Ispra, Italy}


\begin{abstract}
We present a practical approach to inter-compare a range of candidate digital elevation models (DEMs) based on pre-defined criteria and statistically sound ranking approach. The presented approach integrates the randomized complete block design (RCBD) into a novel framework for DEMs comparison.

The method presented provides a flexible, statistically sound and customizable tool for evaluating the quality of any raster - in this case a DEM - by means of a ranking approach, which takes into account a confidence level, and can use both quantitative and qualitative criteria. The users can design their own criteria for the quality evaluation in relation to their specific needs. 

The application of the RCBD method to rank six 1'' global DEMs, considering a wide set of study sites, covering different morphological and landcover settings, highlights the potentialities of the approach. We used a suite of criteria relating to the differences in the elevation, slope, and roughness distributions compared to reference DEMs aggregated from 1-5 m lidar-derived DEMs. Results confirmed significant superiority of CopDEM 1'' and its derivative FABDEM as the overall best 1'' global DEMs. They are slightly better than ALOS, and clearly outperform NASADEM and SRTM, which are in turn much better than ASTER.
\end{abstract}



\end{frontmatter}



\section{Introduction}
Over the past two decades, several Earth observation missions have resulted in finer than 100~m resolution global digital elevation models (DEMs), most of which are shared freely and openly worldwide. These data revolutionized earth sciences and spurred many applications that require accurate information about the shape of Earth’s surface. Consequently, the demand for high quality DEMs across multiple disciplines continues to grow and users around the world are faced with the challenge to understand the DEMs' intrinsic characteristics and to make an informed choice for their particular application.

At this time, at least six different global medium resolution (i.e., 10-100~m) DEMs have been produced using a variety of techniques. We expect more to come in the near future as new technologies and methods are developed. However, most users do not have the resources and expertise to perform an in-depth comparison of different DEMs. Therefore expert advice that provides information pertinent to the user's need will benefit the community to identify the most appropriate dataset. This challenge identified by the Committee on Earth Observation Satellites (CEOS) Working Group on Calibration and Validation (WGCV) was taken up by the Terrain Mapping Subgroup (TMSG) in its Digital Elevation Model Inter-comparison eXercise (DEMIX), by developing open and transparent tools to compare digital elevation models and rank them based on their choice of criteria.

As a major step towards this goal, this paper presents a novel and flexible tool to provide geospatial data users with a practical approach to inter-compare a set of candidate global DEMs based on pre-defined criteria and a statistically sound ranking approach. During the discussions and development, the DEMIX group referred to the method as the `wine contest' as an analogy to a comparison of wines by human judges, but with DEMs instead of wines, and both qualitative and subjective criteria instead of judges. The framework provides the wider geospatial community tailored recommendations regarding available DEM products that are not limited to one domain, geographic area, or landscape type. The method is flexible and customizable in relation to the specific needs and requirements of users in their particular application. We then rank global 1'' DEMs and show that three of them are clearly superior based on the chosen criteria.

\subsection{A brief review of comparing and ranking global DEMs}
The Shuttle Radar Topography Mission (SRTM), which flew in 2000 and released edited DEMs starting in 2004, revolutionized digital topography \cite{Farr2007}. Previously DEMs had been created from stereo photogrammetry, either directly from the imagery or indirectly from the contours on topographic maps, and covered only a single country and were often not freely available \cite{Gesch1999}. SRTM produced and released a globally consistent DEM for latitudes lower than $60^\circ$, initially at 1 arc-second (1'', $\sim30$~m at the equator) grid spacing in the United States and at 3'' ($\sim90$~m) everywhere else, which starting in 2014 was also made available globally at 1'' spacing. The U.S. Geological Survey (USGS) held a workshop in 2005, and the published papers set the standard for assessing SRTM using reference data and determining the quality of the DEM \cite{Gesch2006}. The approaches used either sparsely located control points \cite{Rodriguez2006,Hofton2006}, or comparison to reference DEMs using all points in the grid, comparing the elevations or including derived geomorphometric parameters such as slope \cite{Guth2006}. These workshop papers also emphasized that results vary with different land cover and terrain slope categories \cite{Carabajal2005,Hofton2006}.

The release of the 1'' ASTER global DEM \cite{Abrams2010} led to a large number of papers comparing SRTM and ASTER to each other and reference data, and assessing whether SRTM or ASTER was ``better'' \cite{Reuter2009a,Guth2011,Slater2011}, generally finding SRTM was superior, although at that point the 1'' SRTM was still not available globally. A new milestone was achieved with the  release of the TanDEM-X DEM \cite{Wessel2016,Rizzoli2017}, a global DSM at 0.4'' latitude pixel spacing and 0.4'' to 4'' in longitude. The data set remained limited to scientific use, but produced a number of comparisons to  SRTM and ASTER and ALOS \cite{Grohmann2018}. TanDEM-X DEM is also the basis for the 1'' and 3'' DEMs of the same name and the basis for the commercial WorldDEM and therefore also for the Copernicus DEM \cite{Strobl2020gissq}. Sequential releases of ALOS World 3D \cite{Tadono2016}, NASADEM \cite{Crippen2016}, TanDEM-X DEM, and CopDEM led to additional comparisons of the freely available 1'' DEMs trying to establish which was ``best.'' 

Most studies compare DEMs against each other or with reference data (ICESat/GNSS point elevations or lidar datasets) for a limited number of sites usually located in one country, and considering a single parameter (i.e., elevation differences) and a single criterion, typically based upon popular statistics such as mean, median, standard deviation and root mean square error (RMSE) \cite{Guth2006,Hofton2006,Rodriguez2006,Grohmann2018,Zhang2019,GonzalezMoradas2020,Gdulova2020,Guth2021b,Ebinne2022,GonzalezMoradas2023}. Other criteria include comparisons of contour lines \cite{Grohmann2018,Ebinne2022} drainage network \cite{GonzalezMoradas2020,GonzalezMoradas2023}, and topographic profiles \cite{Grohmann2018,Guth2021b,Ebinne2022}.

In all the cases the ``best'' DEM is the one with the lowest value for the metrics of difference (that is, closest to the reference data), but even with the same criterion, different sites might produce a different ranking. In that case there is no reason to expect that there will be an agreement for the ranking using different criteria. As a result, these studies provide only very limited recommendations of which DEM performs better for a specific use and in a specific region \cite{Florinsky2019,Zhang2019}.%

The availability of 3'' DEMs like the public version of the TanDEM-X DEM and MERIT \cite{Yamazaki2017} produced additional comparisons \cite{Hawker2019}, complicated by the difference in resolution. The free ALOS and CopDEM represent down-sampled versions of commercial DEMs. Commercial DEMs are beyond the scope of this paper because of their cost and the many difficulties comparing DEMs with different spatial resolution.

All of the global 1'' DEMs have gone through sequential editions, starting with the use of ASTER to fill the voids in SRTM, and are all to some degree composite digital surface models \cite{Guth2021a}. The next evolution in DEMs has been the application of machine learning to ``improve'' current global DEMs considering additional external data; \cite{Hawker2022} created the first freely available, global digital terrain model \cite{Guth2021a} named FABDEM. 

While the conceptual difference between a DSM and DTM is clear, the practical matter of creating the DEM from a satellite-based sensor is not so easy. How the sensor integrates the signal over a pixel area, and how much the signal penetrates vegetation, varies with the sensor and the characteristics of the area. \cite{Guth2021a} showed how the global DEM elevations varied within a high resolution lidar point cloud, and \cite{Guth2021b} showed statistical distributions of the DEM elevations above, below, or within a lidar point cloud, and that the results differed among the global 1'' DEMs. Calling these DEMs DSMs is a simplification; current technology does not allow for selection of a DSM or DTM, but simply reports what the sensor measured. Even the derived FABDEM, seeking to adapt CopDEM and create a DTM, cannot create a perfect DTM, and does not improve on CopDEM everywhere. The other DEMs are closer to DSMs but most likely fall somewhere between a real DTM and DSM \cite{Guth2021b}. We compare the global DEMs to both reference DTMs and DSMs; the user must clearly understand the implications of the results, and decide which are valid for their purpose.

\section{Methods}
\label{sec:methods}

Table~\ref{tab:workflow} summarizes the steps to evaluate six global 1'' DEMs with 236 widely distributed test tiles from 24 test areas. We produce a GIS database which contains an evaluations table with numerical metrics comparing the test DEM to a reference DEM, and an opinions table which translates the evaluations to an ordinal ranking that allows for ties and statistical confidence in the results (see Sec.~\ref{sec:wine_contest} for details).

\begin{table*}[ht!]
    \centering
    \caption{The general DEMIX steps as executed for the comparison of global DEMs.}
    \adjustbox{max width=\textwidth}{%
    \begin{tabular}{lllll}
     Step & Section & Software & Task & Details\\
    \toprule
     1   & \ref{sec:test_areas}      & Web downloads     & Obtain high quality reference    &  6 candidate global DEMs, 1'', geographic \\
         &                           &                   & source data and candidate DEMs   &  Source reference DTM: 1-5m UTM projection\\
         &                           &                   &                                  &  Source reference DSM: if available \\ 
    \midrule
     2   & \ref{sec:prep_ref_dems}   & MICRODEM          & Prepare reference DEMs           &  Pixel-is-point  \\
         &                           &                   &                                  &  Pixel-is-area  \\
         &                           &                   &                                  &  For high lat COPDEM, separate Pixel-is-point  \\
         &                           &                   &                                  &  For high lat ALOS, separate Pixel-is-area  \\
    \midrule
     3   & \ref{sec:prep_cand_dems}  & MICRODEM          & Prepare candidate DEMs           &  Convert to ellipsoid, then to geoid  \\ 
         &                           &                   &                                  &  Use grids from national/regional mapping agency  \\
         &                           &                   &                                  &  Use GDAL for areas in the USA \\
    \midrule
     4   & \ref{sec:build_gisdb}     & MICRODEM          & Build GIS Database               &  Compute test area statistics  \\ 
         &                           &                   &                                  &  Create difference outputs  \\ 
         &                           &                   &                                  &  Compute metrics and evaluations table  \\
         &                           &                   &                                  &  Produce opinions table from initial tolerances  \\
    \midrule
     5   & \ref{sec:ranking}         & Jupyter Notebook  & Rank global DEMs                 &  Choose criteria to use  \\
         &                           &                   &                                  &  Adjust tolerances if required \\
         &                           &                   &                                  &  Recreate opinions table  \\
         &                           &                   &                                  &  Filter database for land types or tiles  \\
         &                           &                   &                                  &  Produce final rankings  \\
         &                           &                   &                                  &  Compute confidence levels  \\
         &                           &                   &                                  &  Produce graphics  \\
    \bottomrule
    \end{tabular}
    } 
\label{tab:workflow}
\end{table*}

\subsection{Test areas and DEMIX Tiling}
\label{sec:test_areas_tiles}

\begin{figure}
\centering
\includegraphics[width=0.75\textwidth]{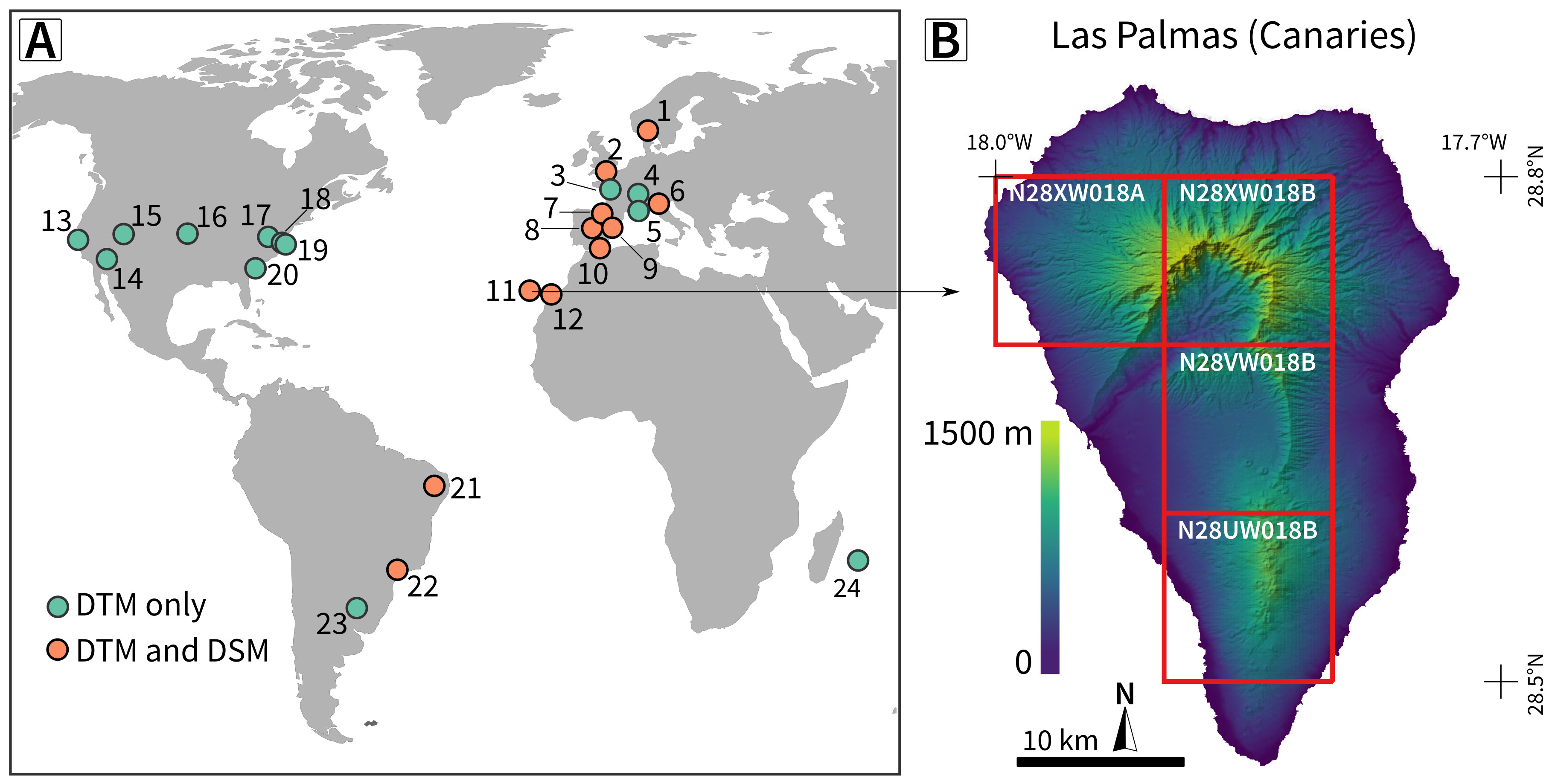}
\caption{A) Location of the 24 test areas made up of 236 DEMIX tiles. B) Distribution of DEMIX tiles over Las Palmas Island. The names of the test areas shown on the map A: 01 -- Norway [9 tiles], 02 -- Oxford [4], 03 -- Caen [6], 04 -- Valonne [9], 05 -- Vanoise [4], 06 -- Trentino [1], 74 -- Pyrenees [2], 08 -- Madrid [35], 09 - Ebro Delta [27], 10 -- Almeria [30], 11 - Las Palmas [4], 12 -- Canary East [18], 13 -- Redwoods [6], 14 -- State Line [13], 15 - Canyon Range [14], 16 - Republican River [19], 17 -- Shenandoah [2], 18 -- Blackwater [6], 19 -- Chincoteague [5], 20 -- Charleston [6], 21 -- Pernambuco [4], 22 - São Paulo [5], 23 -- Uruguay [1], 24 -- La Reunion [6].\label{fig:1_tiles_areas}}
\end{figure}

The reference data for the inter-comparison of global DEMs has significantly higher accuracy elevation values and much smaller sampling spacing than the DEMs to be tested. These reference DEMs were produced by a variety of mapping agencies from around the world with sampling interval between 1--5 m which is more than five times finer spatial resolution than the global DEMs considered. The distribution of the test areas with reference DEMs is shown in Figure~\ref{fig:1_tiles_areas}; some of the reference DEMs provided separate DSMs and DTMs. We decided not to alter the global DEMs beyond converting all to the EGM2008 vertical datum, and only transformed the reference data to match the global DEMs. 

For each of the test areas, one or more DEMIX tiles were extracted. A DEMIX tile covers approximately 10 km x 10 km in size defined on a geographic latitude/longitude grid. This grid covers the entire globe and was developed in the context of DEMIX to allow anyone to use a consistent sampling. The DEMIX tiles provide a global tessellation at a scale fine enough to produce locally significant results, coarse enough to keep the total number of tiles manageable, and nearly equal in size to allow for statistical aggregation and comparative analysis. Technical details, naming convention and grid definition files can be found in \cite{Guth2023}.

The area covered by the reference DEMs varied by location and therefore the number of DEMIX tiles covering each area differs. Reference DEMs do not have to fit perfectly within DEMIX tiles and the expectation is that most will not because of the irregular boundaries of the reference DEMs due to mapping project extent, administrative boundaries, or DEM boundaries in UTM (Universal Transverse Mercator) projection rather than geographic lines. Furthermore, it is not necessary to fill entirely a DEMIX tile, otherwise it would be impossible to include many interesting geographical features such as coastal areas or missing data because some mapping agencies do not edit water and leave voids in lidar-derived DEMs. However, in order to include the tile in our database, we require that at least 75\% of the DEMIX tile be covered by reference and candidate data. Figure~\ref{fig:1_tiles_areas}B shows Las Palmas island area (Location 11). Two of the four tiles are full whereas two tiles covering the coast are $>75\%$ full. The other tiles will not be included because they do not fulfil the 75\% DEMIX sample tile coverage rule. 

Both DTMs and DSMs were used in this evaluation where available. While DTMs were available for all 24 test areas and 236 tiles, DSMs are only available in 11 of those areas (134 tiles) because some mapping agencies do not provide them (Figure~\ref{fig:1_tiles_areas}).

\subsection{Step 1 - Obtain high quality reference source data and candidate DEMs}
\label{sec:test_areas}

For a comparison of global DEMs, the reference DEMs must cover a representative sample of the different landforms observed on Earth. The quality of the global DEMs being compared are not expected to differ across national or continental borders because the same acquisition technique and processing was used to create each of the DEMs under consideration. Therefore, it was necessary to be sure that a representative sample was available and covered landforms available on all continents. As a first approximation, this was achieved by calculating statistics and classifying landforms found within sample tiles. Summary statistics were calculated across all 236 tiles using the reference 1'' DTMs. The results are summarised in Table~\ref{tab:reference_dem_stats} for the following characteristics: 

\begin{itemize}
 \item Mean elevation in meters;
 \item Mean slope in percentage \cite{Evans1980};
 \item Mean roughness in percent, with roughness defined as the standard deviation of the slope in a 5x5 window \cite{Grohmann2011};
 \item Relief -- elevation range in meters;
 \item Percent forest -- percentage that is in a forest category \cite{Buchhorn2020}; 
 \item Percent urban -- percentage that is urban \cite{Buchhorn2020};
 \item Percent barren -- percentage that is one of the barren categories \cite{Buchhorn2020}.
\end{itemize}

The results show that a wide range of surface characteristics are captured in the test tiles. These summary statistics are further supported by land cover and geomorphometric classifications.

\begin{table*}[ht!]
    \centering
    \caption{Reference DTM collection summary statistics for all 236 DEMIX tiles.}
    \adjustbox{max width=\textwidth}{%
    \begin{tabular}{lccccccc}
     & Average & Average & Average & Relief & Forest & Urban & Barren \\
     & elevation (m) & slope (\%) & roughness (\%) & (m) & in tile (\%) & in tile (\%) & in tile (\%)  \\
    \toprule
     Maximum   & 2226.1 & 76.7 & 22.7 & 2363.8 & 99.7 & 96.9 & 98.7 \\ 
     Mean      &  564.9 & 15.4 &  4.8 &  530.8 & 29.3 &  6.7 & 22.3 \\ 
     Median    &  461.5 & 10.4 &  3.6 &  356.1 & 12.2 &  1.2 &  8.2 \\ 
     Minimum   &   -0.9 &  0.1 &  0.0 &    2.8 &  0.0 &  0.0 &  0.0 \\ 
     Std Dev   &  527.7 & 14.3 &  4.0 &  516.1 & 32.3 & 15.2 & 27.9 \\ 
    \bottomrule
    \end{tabular}
    } 

\label{tab:reference_dem_stats}
\end{table*}

\begin{figure*}
\centering
\includegraphics[width=0.99\textwidth]{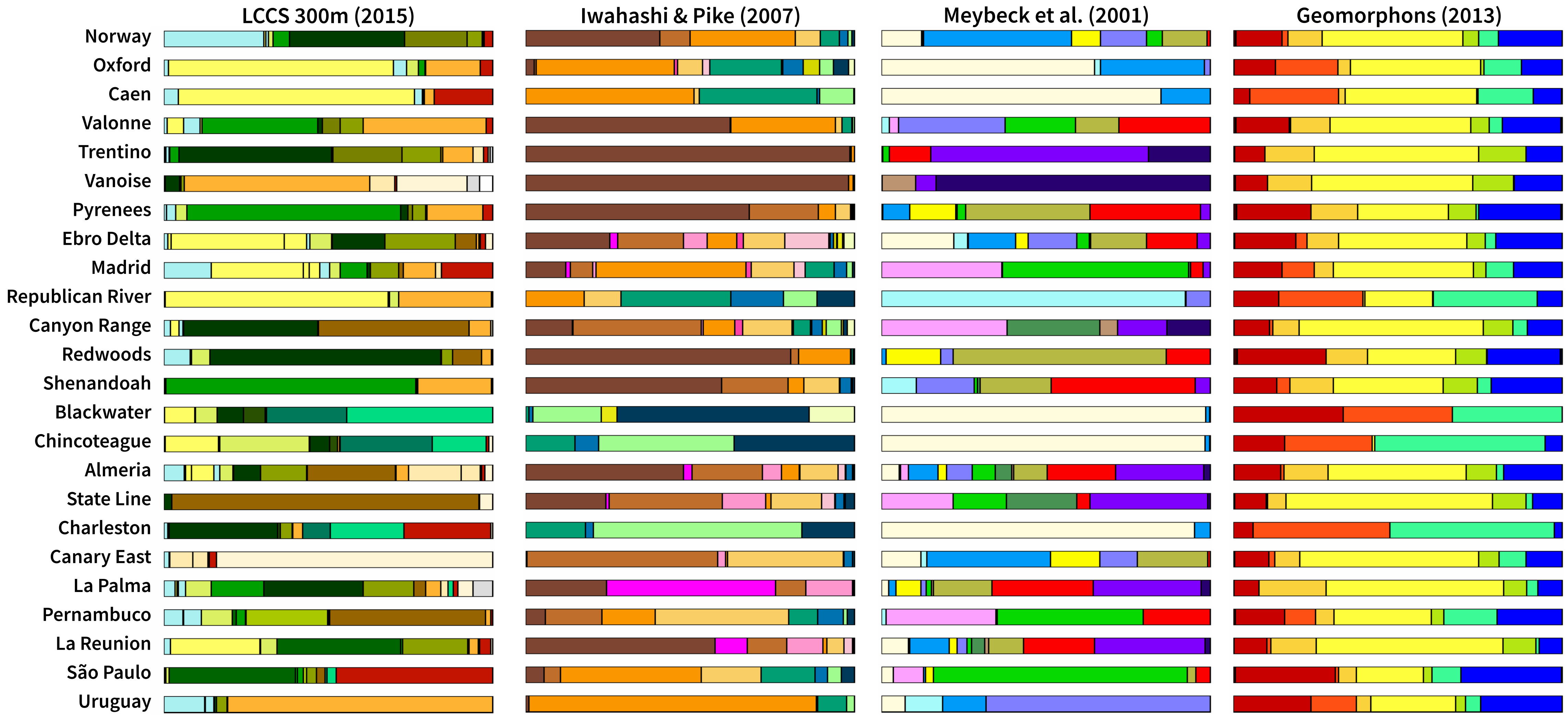}
\caption{Tile classifications for the 24 test areas in our sample, for one landcover classification and three geomorphometric landform classifications. Areas are arranged from north to south. Legend for colors is available in Supplementary Figure 1.\label{fig:2_tiles_class}}
\end{figure*}

Figure~\ref{fig:2_tiles_class} presents the classification associated with each of the 24 test areas. The left column displays the land cover classes using the United Nations Land Cover Classification System with 38 categories \cite{Buchhorn2020}. The colour denotes the class whereas the width of the bar denotes the overall land cover percentage of that particular cover class. The right three columns of Figure~\ref{fig:2_tiles_class} present the geomorphometric classifications from left to right: \cite{Iwahashi2007}, \cite{Meybeck2001}, and geomorphons \cite{Jasiewicz2013}. These geomorphometric classes range between 7 to 16 categories each. Furthermore, Figure~\ref{fig:2_tiles_class} reveals the challenge in trying to neatly characterize a 100 km\textsuperscript{2} tile because no two DEMIX tiles, even adjacent ones, are alike from their surface characteristics. Our tile selection clearly represents a large range of different landscapes and provides a much better diversity of surface characteristics than previous efforts comparing global DEMs.

\subsection{Step 2 - Prepare reference DEMs}
\label{sec:prep_ref_dems}

In order for the reference DEMs to be compared directly with the global DEMs they must be transformed into the same spatial sampling (grid) and vertical reference (datum). The following key global DEM characteristics were managed to produce the reference DEMs. The pixels in the geographic DEMs are rectangular (larger in the north-south direction once away from the equator), and rotated slightly compared to the UTM projection of the source DEM.

\textit{Match Grid Type} -- the global DEMs under consideration have different grid representations which affect their spatial location: pixel-is-point and pixel-is-area \cite{Guth2021a}. Only one quarter of the area is shared by pixels defined by the two representations, and although they have the same nominal coordinates in uneven terrain they are unlikely to have the same elevation.

\textit{Spatial Sampling} -- a 1'' DEM cannot be compared directly with a reference DEM having a spatial resolution of 1-5~m because of the significant differences in spatial scale. The global DEMs have a spatial sampling of approximately 30~m (1'') at the equator and their sensing techniques can be assumed to deliver an average elevation in each given cell. Therefore, the reference DEMs must be aggregated in order to match the spatial sampling of the global DEM.

\textit{Aggregation} -- global DEMs are provided in geographic coordinates whereas reference DEMs typically use local projections such as UTM to avoid spatial distortions. While interpolation has been used to solve the reprojection issues, \cite{Guth2021} showed that re-interpolation changed the characteristics of the DEM and that care in the algorithms allows accurate computations with the geographic coordinates without introducing significant interpolation errors. The projected coordinates of each pixel in the source reference DEMs are converted to WGS84 latitude and longitude, and mapped to the corresponding pixel in the test DEM. The value for the reference 1'' DEM is the mean of the points in the high resolution source DEM, ranging from about 36 points for a 5~m DEM to about 900 for a 1~m DEM.

\textit{Latitudinal Effects} -- the pixel spacing changes with latitude for CopDEM and ALOS  but remains constant for the others. 

\textit{Vertical Datum} -- an elevation model is referenced to a vertical datum. We moved the reference DEMs to the EGM2008 vertical datum using bilinear interpolation and the highest resolution available transformation grids from the PROJ library (\url{https://proj.org/}) or using GDAL to provide combined vertical and horizontal datum shifts for the United States DEMs. Both the PROJ library and GDAL came via OSGeo4W install, (\url{https://www.osgeo.org/projects/osgeo4w/} downloaded 11 April 2023). In a few cases we used a local transformation grid, not included in the PROJ library and which does not have an EPSG code.

The DEM characteristics were carefully monitored during the preparation of the reference DEMs. This was achieved by using MICRODEM version 2023.6.20 \cite{Guth2009,Guth2022md} as the reference software for this experiment because it allowed access to the source code to analyse and fix any issues with the preparation of the reference DEMs and/or transformations of the global DEMs. The authors are not aware of other software that will do the aggregations correctly from UTM to geographic grids, in both pixel-is-area and pixel-is-point geometries with the same simplicity and transparency.

For each source DTM, the following reference DTMs were produced to cover the different global DEMs in the test areas:

\begin{enumerate}
\item 1'' x 1'' pixel-is-point DTM;
\item 1'' x 1'' pixel-is-area DTM;
\item 1'' x CopDEM spacing pixel-is-point for high latitude DEMs;
\item 1'' x ALOS spacing pixel-is-area for high latitude DEMs;
\end{enumerate}

If the reference DEM also had a DSM, the same types of reference DSMs were produced. Up to eight reference DEMs could be required covering each DEMIX tile. This makes it possible to perform quantitative analysis between reference DEMs and the global DEMs without the need for further interpolation or adjustment of the candidate DEM.

\subsection{Step 4 - Prepare candidate DEMs}
\label{sec:prep_cand_dems}

We moved NASADEM, ALOS, SRTM, and ASTER from the EGM96 to the EGM2008 vertical datum, forcing them to be of floating point representation, using the 15 arc-minute grid for EGM96 (the best available) and the 1 arc-minute grid for EGM2008. CopDEM and FABDEM are delivered in floating point precision for elevation with EGM2008 and did not require any changes.

\subsection{Build the GIS Database}
\label{sec:build_gisdb}
Creating the GIS database used for the rankings involves a number of design consideration, including the selection of criteria and the tolerances to decide if small numerical differences represent significant differences.

\subsubsection{Criteria used for DEM Comparison}
\label{sec:criteria}

For this paper, a pixel-by-pixel comparison was applied for the assessment of global DEMs, using a reference DEM with higher accuracy elevation values. A pixel-by-pixel comparison computes the difference between a parameter calculated using the global DEM and the reference DEM; the parameter can be either the elevation or a derived focal parameter. The reference DEM (DTM and DSM if available), is assumed to be the best representation of elevation and derived parameters for a particular sample location. Consequently, any differences will be considered as the error of the candidate global DEM.

The quantitative assessments carried out on the global DEMs were based on pixel-by-pixel differences of three geomorphometric parameters: elevation, slope, and a roughness index. The following parameters were computed for the 6 candidate DEMs for all DEMIX tiles with respect to the reference DTM and reference DSM where available: elevation differences (ELVD), slope differences (SLPD) using the \cite{Evans1980} algorithm, and roughness differences (RUFD) from the standard deviation of slope in a 5x5 window \cite{Grohmann2011}. 

These parameters were chosen because they measure different aspects of the DEM and the underlying terrain, and are not closely correlated. As derivatives of elevation, the slope and roughness parameters are more sensitive to non-systematic DEM elevation errors and are also among the most used key parameters applied in many analyses for the characterisation of DEMs. Among the various roughness indices \cite{Grohmann2011}, the standard deviation of slope can be considered a flow-directional roughness index \cite{Trevisani2016} being computed in the direction of the gradient.

\begin{figure*}
\centering
\includegraphics[width=0.95\textwidth]{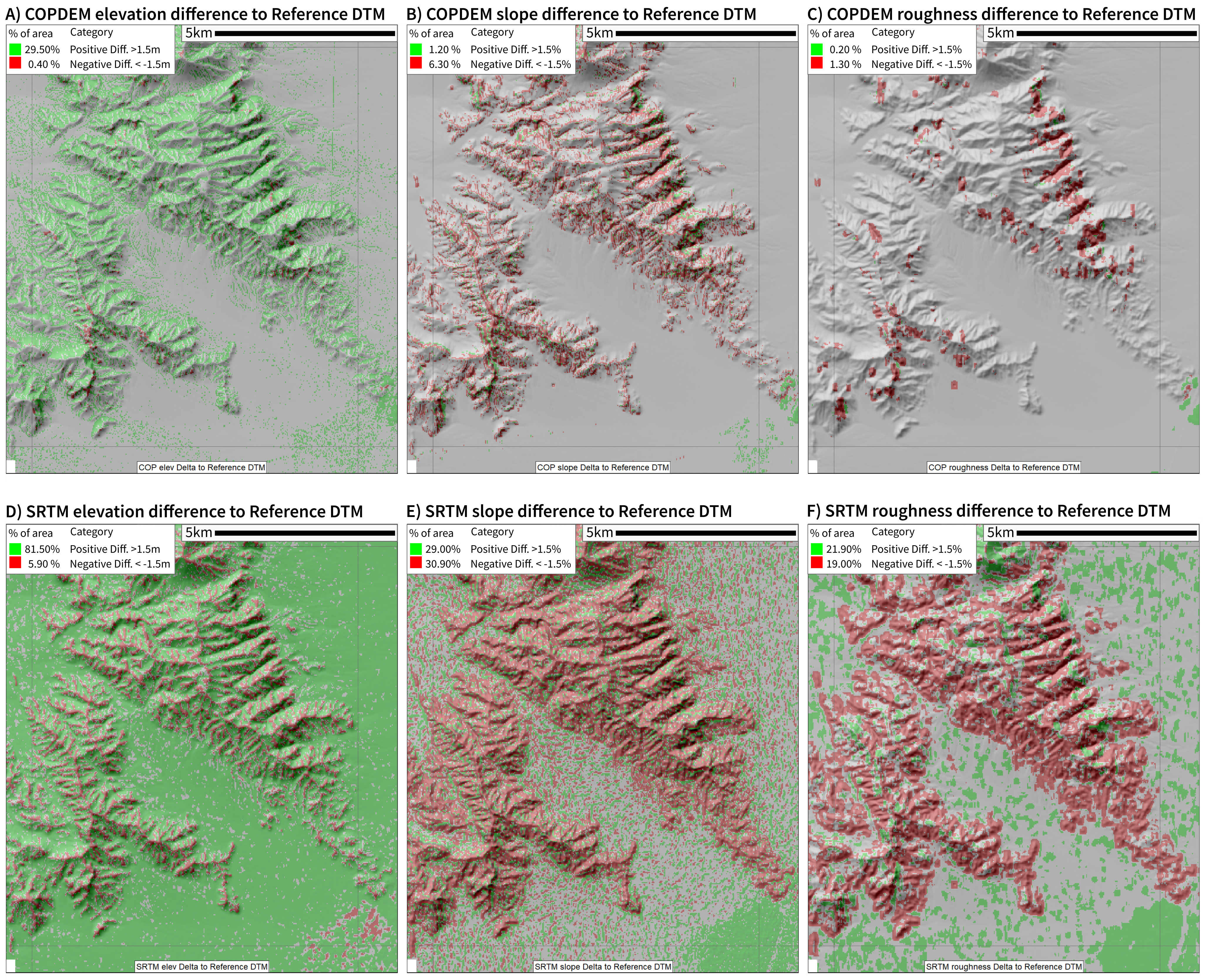}
\caption{Difference maps for Tile N35VW116G for CopDEM and SRTM compared to the reference DTM. Difference Maps for all six Global DEMs are presented in Supplementary Figure 2.\label{fig:3_shaded_differences}}
\end{figure*}

\begin{figure*}
\centering
\includegraphics[width=0.95\textwidth]{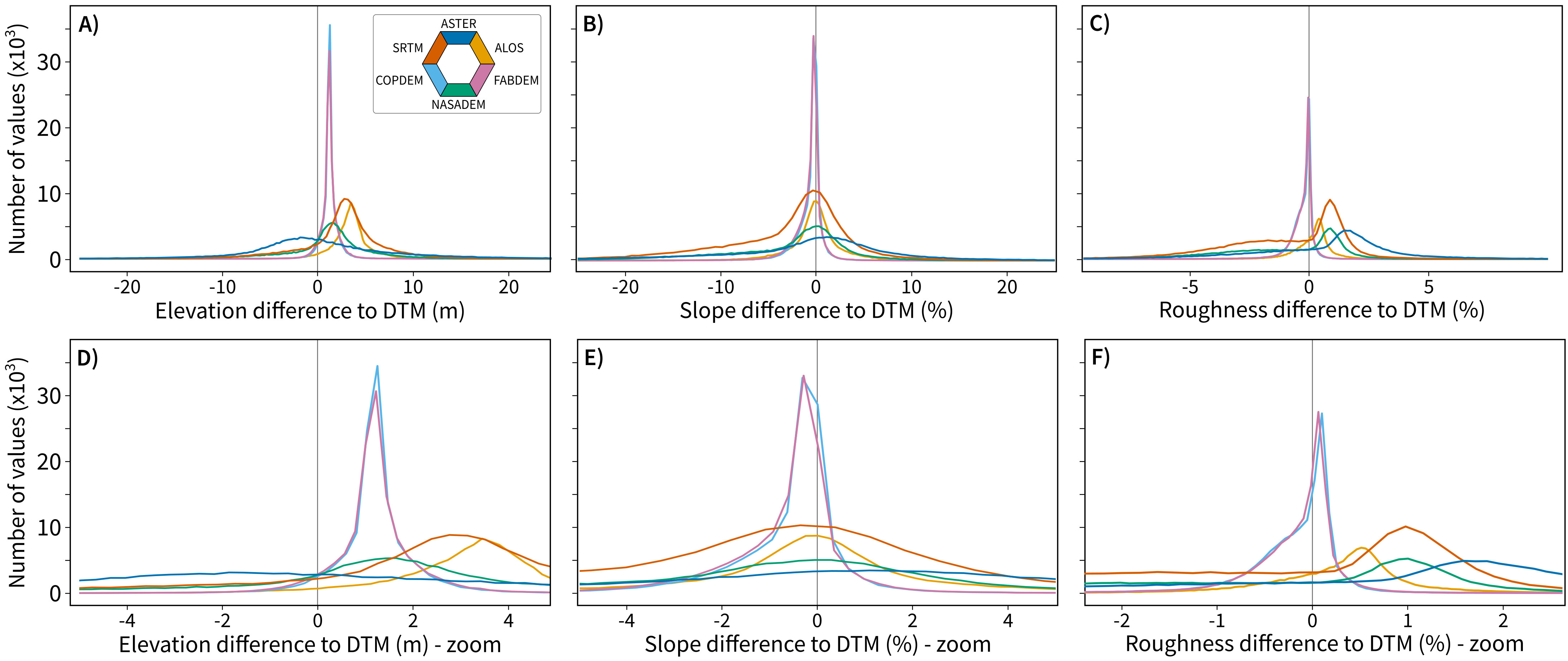}
\caption{The distribution of elevation, slope and roughness differences for tile N35VW116G, compared to the reference DTM. Positive values mean the candidate DEM has higher values than the reference. The top figures show the entire distribution, and the bottom highlights the region around zero.\label{fig:4_histograms_stateline}}
\end{figure*}

The differences between the reference DEM and global DEM were always computed in the same manner:

\begin{equation*}
difference = global_{DEM} - reference_{DEM}
\end{equation*}

where positive values indicate that the global DEM has a higher value than the reference DEM at a specific pixel. The elevation difference parameter cannot be ranked because the range of outputs will be signed and ideally distributed around zero if the DEM is unbiased. Figure~\ref{fig:3_shaded_differences} shows the difference maps for tile N35VW116G for CopDEM and SRTM, draped on a hillshade. Figure~\ref{fig:4_histograms_stateline} shows the difference distributions for this tile, for all 6 DEMs. In this tile there is a slight positive bias for elevation, and negative bias for slope for CopDEM, which is a variable result among all the tiles we considered. The very well defined modes close to 0 occur in almost all the 236 tiles.

In order to achieve a quantitative assessment required for the randomized complete block design (RCBD) method based on the above parameters, it was necessary to compute derived statistics from the distribution of the differences for each of the 236 DEMIX tiles:

\begin{itemize}
\item{Standard deviation (STD)} -- a characteristic of the difference distribution, which is not related to the tolerance we use to create ties; 
\item{Average deviation (AVD)} -- the mean difference was subtracted from each difference, and the average of its absolute value computed \cite{Press2007};
\item{Root mean square error (RMSE)};      
\item{Mean absolute error (MAE)};
\item{Linear error 90 (LE90)} -- the 90\textsuperscript{th} percentile of the distribution of the absolute differences.
\end{itemize}

For all of the chosen metrics, the quantitative assessment rule is that lower is better. These metrics output only non-negative values which allows them to be easily ranked. These metrics measure different properties of the difference distribution, putting differing weights on the tails of the distribution. The database also includes the mean and median for each difference distribution, because they provide additional context in interpreting the results, despite not being useful for ranking purposes.

The three parameters and five metrics for each DEMIX tile gives 15 different criteria on which to apply the RCBD methodology. These are presented in the following manner: ELVD\_RMSE, where the first acronym specifies the computed parameter, and the second acronym specifies the metric applied.

\subsubsection{Per Pixel Land Types}
\label{sec:land_types}

Global DEMs perform differently depending on the underlying characteristics of the surface. As shown in Figure~\ref{fig:2_tiles_class}, any single DEMIX tile incorporates a variety of surface characteristics whose tendencies can be lost when looking at overall metrics. In addition to computing statistics for every pixel in the tile, we computed separate pixel-by-pixel statistics for different slope and land cover classes. We used three land cover classes (forest, urban, and barren) from the Copernicus Global Land Cover Layers -- Collection 2 \cite{Buchhorn2020}, and four slope categories Cliff (slope $>$ 50\%), Steep (slope $>$ 25\% and $<$ 50\%), Gentle (slope $<$ 25\% and $>$ 12.5\%), and Flat (slope $<$ 12.5\%).

\subsubsection{The DEMIX Database}
\label{sec:demix_db}

MICRODEM \cite{Guth2009,Guth2022md} produced the DEMIX GIS database \cite{Guth2023_dbv2}, combining the tile characteristics, an evaluations table, and an opinions table with an initial set of tolerances (see Section~\ref{sec:wine_contest}). The evaluations table contains the numerical results for each criterion for every tile, and will not change during the ranking. Calculating the evaluation table and tile characteristics are a relatively slow processes, but will not have to be recomputed unless the DEMs or criteria change. It can be expanded any time by additional tiles or criteria.

The opinions table ranks the six DEMs from 1 to 6 (low score is best), with the possibility of ties. We assign a tolerance for each criterion; evaluations within the tolerance will be treated as a tie. Many of the candidate global DEMs are rounded to integer type, therefore it does not make sense to treat differences on the order of centimeters or even decimeters as significant for any elevation-related metric. Based on our experience with global DEMs, we have chosen the tolerances below; for other applications, different tolerances might be appropriate. For the elevation parameters we chose a tolerance of 0.5~m. For slope, an elevation error of 0.5~m over a 30~m horizontal distance would lead to a slope error of about 1\% and we picked a tolerance of 0.5\%. We used a tolerance of 0.2\% for the roughness. Given the evaluations table and the tolerances, the opinions table can be created on the fly very rapidly after a change of the tolerances or filter of the database. Output from the database can be extracted in an interchangeable, human readable, comma separated values (.csv) file format.


The current DEMIX database \cite{Guth2023_dbv2} includes 55,699 opinions, which are rows found within the database. The following fields are included to help reading the database:

\begin{itemize}
\item Tile name, test area, and centroid coordinates;
\item Tile characteristics required for per tile filters; 
\item Whether the reference DEM is a DTM or a DSM (\ref{sec:test_areas});
\item The land type, with ALL for every pixel, or one of the seven categories defined above if there are at least 100 pixels of that type within the tile, and the percentage of the tile in the land type;
\item Criterion (quantitative assessment metric) name for the 15 applied metrics (plus six other signed metrics for analysis purposes not applicable to the RCBD method) defined above;
\item Evaluations table using all six candidate DEMs;
\item Tolerance used to populate the opinions table and identify ties;
\item Opinions table for each computed metric including the identification of potential ties due to the pre-defined tolerances.
\end{itemize}

A summary of the composition of the DEMIX database used for the ranking of the candidate global DEMs is presented in Table~\ref{tab:demix_db}. The table shows the number of tiles with more than 100 pixels (out of roughly 130,000 pixels) within a tile based on the eight defined surface types. This adds up to the total number of opinions available in the database to produce the final ranking. For example, of the 236 reference DTM tiles (the ``ALL'' surface type), only 152 have minimum required 100 `cliff' surface type pixels whereas 181 tiles have the minimum required number of `urban' surface type pixels.

Any combination of opinions can be used in the final ranking. The database can be filtered to show only opinions for a single land type, like `steep' or `barren,' or a user may only be interested in the tiles where the forest cover type covers at least 50\%--75\% of the tile, or where the relief is greater than 1000~m. This would filter the evaluations table to only include tiles that fulfil the desired requirement.

The database stores all the needed outputs, including the evaluations and opinions tables, for the RCBD method applied to the comparison of the global DEMs. Consequently, they do not need to be recomputed when a subset is used to compute the final ranking.

\begin{table*}[ht!]
    \centering
    \caption{Composition of the DEMIX database in terms of Land Types.}
    \adjustbox{max width=\textwidth}{%
    \begin{tabular}{lccccccccccc}
    \cmidrule{3-10}
        &    & \multicolumn{8}{|c|}{Land Types}\\
    \midrule
     &  &  &  &  &  &  &  &  &  & Total Types & Total \\
    REF\_TYPE  & AREA & ALL & BARREN & FOREST & URBAN & FLAT & GENTLE & STEEP & CLIFF & Present & Opinions \\
    \midrule
    DSM &  11 & 134 & 129 & 116 & 121 & 134 & 132 & 131 & 108 & 897 &  15080 \\ 
    DTM &  24 & 236 & 231 & 201 & 181 & 236 & 219 & 190 & 152 & 1494 & 24705 \\ 
 Total  &  24 & 370 & 360 & 317 & 302 & 370 & 351 & 321 & 260 & 2391 & 39785 \\ 
    \bottomrule
        &    &     &     &     &    &    &    &     &     &     & 20640
    \end{tabular}
    } 
\label{tab:demix_db}
\end{table*}

\subsection{Rank Global DEMs}
\label{sec:ranking}
We implemented statistical tests to use the GIS database to rank the DEMs and provide statistical significance.  A Jupyter notebook performs the calculations and creates graphical output.

\subsubsection{Choice of Criteria}
\label{criteria}
This step is one of the most challenging parts of the method because it relates to the choice and calculation of the quantitative criteria, which in the end will be the basis for the ranking of the global DEMs. For this study, the chosen metrics are based on the computed differences to the reference DEMs. The following will underpin the choice of the quantitative assessments for this DEMIX ranking.

\begin{figure}[!h]
\centering
\includegraphics[width=0.9\textwidth]{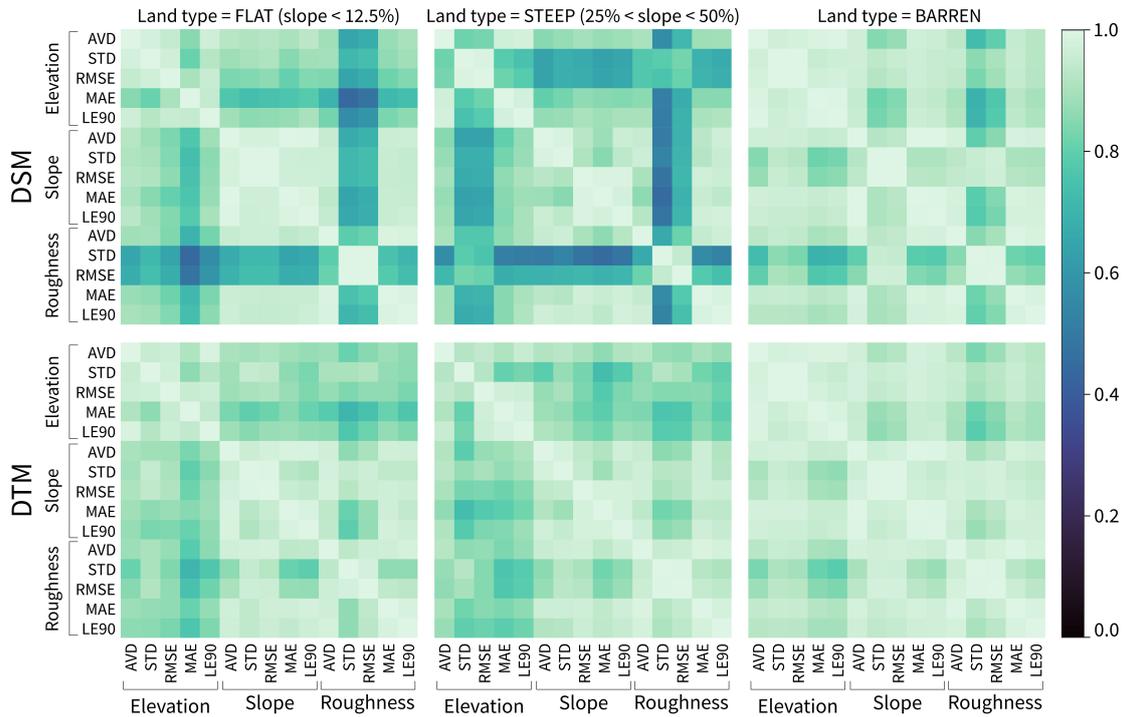}
\caption{Correlation matrix among the 15 criteria, for 236 DEMIX tiles using reference DTMs and DSMs for 3 different land types. Note the generally high correlations, particularly among the five roughness criteria and the five slope criteria. The lowest correlations are between the elevation and roughness criteria. Matrices computed using the Pearson correlation coefficient.\label{fig:5_corr_matrix}}
\end{figure}

First, it is advantageous to use metrics that are not highly correlated because then the amount of diverse information is maximized for the comparison. To illustrate this, correlation matrices were computed based on the 15 quantitative assessment metrics presented for all 236 DEMIX tiles using CopDEM and the reference DTM and DSM where available (Figure~\ref{fig:5_corr_matrix} -- leftmost column matrices) for the land types flat, steep, and barren. The five metrics appear to be correlated within each parameter but less correlated between parameters. For example, generally high correlations (light green) are observed particularly among the five roughness metrics and the five slope metrics. The lowest correlations (darker blue) observed are between the elevation and roughness parameters.

\clearpage

Second, although the design of the RCBD method precludes the use of quantitative magnitudes that cannot be ranked (e.g., the mean) it is of interest to explore magnitudes that provide insights into the comparison of global DEMs. They provide valuable information to interpret the outcomes, as shown in Figure~\ref{fig:4_histograms_stateline}, where the distribution of pixel differences between the reference DSM and DTM are presented for a single tile. Such information provides insights into whether a candidate global DEM is higher or lower, steeper or gentler, rougher or smoother compared to the reference DEM.

\subsubsection{Statistical Confidence with Mixed Quantitative and Qualitative Opinions}
\label{sec:wine_contest}

Ranking a collection of DEMs from a given set of candidates leads to a conceptually similar problem to comparing wines. We use this analogy in order to develop a novel, extensible and user-tuneable comparison of global DEMs. Since it is the object to be ranked, we identify DEMs with wines. For every combination of DEMIX tile and criterion there will be one ranking, formally similar to the one produced by a judge (tile and criterion) analyzing wines (DEMs). We are aware that for wines, most criteria are subjective and require considerations over many sensory aspects. However, objective opinions can be addressed as well. We wish to underline that the method can also be applied to comparisons of other geospatial products and is in no way limited to ranking of global DEMs, and only requires that the products be ranked according to defined criteria.

The proposed DEMIX RCBD ranking provides a statistical procedure to use the results from quantitative and/or qualitative criteria to rank DEMs and provide statistical confidence to the final ranking. The adopted statistical procedure is formally known as the randomized complete block design \cite{Addelman1969}. We apply the RCBD in the context of an inter-comparison of global 1'' DEMs to identify the most appropriate DEM for a given set of criteria and sample sites, while providing statistical confidence to the outcomes. We use multiple criteria, which can be either qualitative or quantitative.

Based on this design (RCBD) two tables will be built. The first one will be denoted here as the evaluation table (Table~\ref{tab:wine_contest_example}, left side). It has $k$ columns (one per global DEM) and $N$ rows (one for each criterion applied to each site), holding the numerical results from evaluating the criterion for a site and for a DEM.  For example, one row might hold the elevation RMSE values for a particular site. A ranking for each row of the evaluations table is produced and stored as the corresponding row of the opinions table (Table~\ref{tab:wine_contest_example}, right side). Each row of the opinion table holds a ranking, which (in a case without ties), is a permutation of the integer values from 1 to $k$. The opinion table can be interpreted as the starting point of the RCBD, so no requirements are posed over how it was produced. Thus the evaluations can be either qualitative (i.e., subjective based on the opinion of an expert) or quantitative based on some objective method that can produce a numerical result amenable to be sorted (like the RMSE). If a criterion cannot produce a ranking it cannot be part of the RCBD method. The ranking is built with the rule of ``lower is better.'' In the RCBD context, the following associations can be made:

\begin{itemize}
\item The DEMs are evaluated in an arbitrary order and therefore the test is randomized. This is important only for subjective criteria.

\item There must be a valid opinion for every wine based on their individual taste, so the test is complete. There are no voids in the opinions table.

\item There is no reason to expect the criteria (block) to produce identical opinions for the same set of wines.
\end{itemize}

\begin{table*}[ht!]
    \addtolength{\tabcolsep}{-0.4em}
    \centering
    \caption{A realistic example applied to the inter-comparison of DEMs -- the evaluations table (left side) records the assessment outcomes for each of the candidate DEMs. The opinions table (right side) translates the evaluations table to ranked opinions. ELVD: elevation difference from reference. SLPD: slope difference from reference. RUFD: roughness difference from reference. RMSE: Root mean square error. LE90: Linear error at 90\%. MAE: Mean absolute error. AVD: Average deviation.}
\label{tab:wine_contest_example}
    \adjustbox{max width=\textwidth}{%
    \begin{tabular}{lcccccccccccccc}
    \quad & \multicolumn{6}{c}{Evaluations} & \quad & \multicolumn{6}{c}{Opinions} & \quad \\
    \cmidrule(lr){2-7} \cmidrule(lr){9-14}
    Criterion    & FABDEM & CopDEM & ALOS & NASADEM & SRTM & ASTER   & tolerance &   FABDEM & CopDEM & ALOS  & NASADEM & SRTM & ASTER   & \quad \\
    \cmidrule(lr){2-7} \cmidrule(lr){9-14} 
     ELVD\_RMSE (m)  &  2.46  & 2.44   & 4.42 & 5.36    & 5.78 & 9.00    & 0.5       &   1.5    & 1.5    & 3     & 4.5     & 4.5   &  6      & \quad \\  
     ELVD\_LE90 (m)  &  2.38  & 2.29   & 6.04 & 8.36    & 8.67 & 14.20   & 0.5       &   1.5    & 1.5    & 3     & 4.5     & 4.5   &  6      & \quad \\  
     SLPD\_RMSE (\%) &  3.52  & 3.47   & 3.79 & 9.22    & 9.04 & 12.48   & 0.5       &   2      & 2      & 2     & 4.5     & 4.5   &  6      & \quad \\  
     SLPD\_MAE  (\%) &  1.28  & 1.19   & 2.20 & 5.83    & 5.70 & 8.26    & 0.5       &   1.5    & 1.5    & 3     & 4.5     & 4.5   &  6      & \quad \\  
     RUFD\_AVD  (\%) &  0.51  & 0.52   & 0.63 & 2.28    & 2.22 & 2.61    & 0.2       &   2      & 2      & 2     & 4.5     & 4.5   &  6      & \quad \\  
     RUFD\_RMSE (\%) &  1.24  & 1.25   & 1.03 & 3.15    & 3.09 & 3.70    & 0.2       &   2.5    & 2.5    & 1     & 4.5     & 4.5   &  6      & \quad \\  
    \cmidrule(lr){2-7} \cmidrule(lr){9-14}  
    \quad        &        &        &      &         &      &         & $Ri$      & 11.0     & 11.0   & 14.0  & 27.0    & 27.0  & 36.0    & \quad \\
    \quad        &        &        &      &         &      &         & $Rj^2$    & 121.0    & 121.0  & 196.0 & 729.0   & 729.0 & 1296.0  & sum=3192.0 \\
    \end{tabular}
    } 
\end{table*}

The null hypothesis is that there is no difference among the DEMs, and a consensus based on the opinions can not be achieved. However, if the null hypothesis is rejected, then the contest ranking is not based on chance (given a chosen confidence level) and some conclusions can be obtained.

Criteria that in general cannot be ranked, i.e., whose results cannot be interpreted in the sense of better or worse, are not applicable. Ties, individual results which are considered equally good (or equally bad), can be considered by applying the mid-rank procedure \cite{Amerise2015} and arise after considering tolerances.

This example illustrates some salient features of the procedure that can help users compare DEMs:

\begin{itemize}
\item The evaluations table can hold both qualitative and quantitative evaluations; qualitative ones must be presented in such a manner that allows one to rank them accordingly; 
\item Quantitative outcomes recorded in the evaluations table are translated to the opinions table according to the mid-rank procedure;
\item Numerical tolerances should be included in the application of quantitative criteria which reflect the uncertainty of the resulting values. 
\end{itemize}

When applicable, tolerances should account for measurement uncertainty to identify minor differences in the evaluations table that will have an impact in the opinions table but are not really different. For example, results in the evaluations table are recorded to two decimal places, but many of these DEMs only record elevation to the nearest meter and thus differences of centimeters or even decimeters would not imply a significant difference among DEMs. Tolerances can also reflect chosen reference data comparison algorithms, and thematic requirements when justified to support the creation of the opinions table.

With the opinions table in-hand, the overall opinion ranking can be computed where $R_j$ is the sum of the ranked value for a given DEM. Lower is better, since the rank order is from 1=best to $k$=worst. Sorting the $R_j$ values might produce the requested ranking result. However, there is yet no statistical confidence associated with these rankings, so the next step is to apply the appropriate statistical tests.

\subsubsection{Integrating Statistical Confidence to the Ranking}
\label{sec:statistical_confidence}

A statistical confidence must be associated to the final rankings because it is imperative that the outcomes produced are not due entirely to chance (as might happen if the opinions table rankings are taken naively). 

One of the strengths of the RCBD method is that both quantitative and qualitative evaluations can be integrated into the contest. In addition, quantitative evaluations using different criteria need not be commensurate, thus precluding the ANOVA approach. Therefore, a non-parametric test should be used, which can operate over both types of rank results. The chosen non-parametric test is the Friedman Test \cite{Friedman1937}, because among the non-parametric choices at hand, it is the best alternative for minimizing the risk of paradoxical results \cite{Fey2012}. The general Friedman's statistic $\chi_{r}^2$, is presented in Eq.~\ref{eq:friedman}, valid with or without ties ($N$= number of opinions, $k$=number of DEMs).

\begin{equation}
{\chi_{r}^2} = \dfrac{N(k-1)\left [ \sum\limits_{j=1}^{k}\dfrac{R_{j}^2}{N}-C_{F} \right ]}{\sum\limits_{i=1}^{N}\sum\limits_{j=1}^{k}{r_{ij}^2} - C_{F}}  ; C_{F}=\dfrac{Nk(k+1)^2}{4}
\label{eq:friedman}
\end{equation}

The entries in the opinions table (Table~\ref{tab:wine_contest_example} -- right side) are denoted as the elements $r_{ij}$, and the final row presents the column sums by $R_j$. The remaining values to compute the Friedman statistic ($\chi_{r}^2$) can be extracted from the same table: with $C_F=441$, sum of ${r_{ij}}^2=537.0$, and sum of ${R_j}^2=3192.0$, then $\chi_{r}^2=28.871$.

For a given $k$, $N$ and confidence level alpha, the Friedman's statistic value $\chi_{r}^2$ is compared to the critical value $\chi_{r_crit}$; if ${\chi_r}^2$ is larger, the null hypothesis is rejected, and the conclusion is that we cannot accept that the DEMs are equivalent. Since we are willing to accept ties, the standard critical values tables from the Friedman Test are not suitable. From the table for $k=6$ provided by \cite{LopezVazquez2021} and \cite{LopezVazquez2020}, and at the 95\% confidence level, the row $N=6$ offers a critical value of $\chi_{crit}=10.489$. Compared to the $\chi_r$ result of $28.871$ which is larger, one should reject the null hypothesis that the opinions table entries are purely at random, implying that the DEMs are not equivalent.

The null hypothesis has been rejected based on the Friedman Test, so there are statistically significant differences among the set of DEMs under consideration. The lower the values of $R_j$, the better, but it is still necessary to assess whether the difference between the pairs of ranked candidate DEMs are statistically significant, or otherwise conclude that the pair is tied. The process is denoted as post-hoc analysis and there exist different options to carry out such an analysis. In this case and following \cite{Pereira2014}, we propose to use the test by \cite{Dunn1961} applying the Bonferroni correction. 

A pair of DEMs is considered significantly different if 

\begin{equation}
| R_i - R_j| \geq z_{1 - \alpha/k/(k-1)}\sqrt{\dfrac{Nk(k+1)}{6}} 
\label{eq:bonferroni-dunn}
\end{equation}

where $z$ stands for the inverse of the cumulative normal distribution, and $\alpha$ is the prescribed confidence level.

\subsubsection{Step 5 - Jupyter notebook implementation}
\label{sec:notebook}

The final step ranks the candidate global DEMs. This ranking is based on statistical computations just described in order to provide confidence in the obtained result.

The needed functionality to read the database and implement the statistical procedures required to produce the final rankings was made available through a Jupyter Notebook \cite{Kluyver2016} using Python version 3.10.x and the Pandas, Numpy, Matplotlib, and Seaborn libraries \cite{Harris2020,Hunter2007,McKinney2011,PSF2021}.

The DEMIX Jupyter notebook can be run on a user's local machine or in the cloud using the Google Colab platform \cite{Grohmann2023_iasi_notebook}. As input this Jupyter notebook takes the GIS database produced by MICRODEM ignoring the signed mean and median statistics. 

The information provided in the database allows the user to directly select opinions most appropriate to their requirements. For each run, the Jupyter notebook computes the final rankings based on the chosen opinions including the required confidence levels to support the outcomes. Furthermore, the DEMIX Jupyter notebook provides tools to analyse outputs by creating graphics and figures to help understand the final rankings. Some results are highlighted in the next section.


\section{Results: Applying the RCBD to Compare Global DEMs}
\label{sec:results}

One goal of the DEMIX RCBD method is to provide DEM users with a procedure that can be applied anywhere for the inter-comparison of global DEMs so that a choice can be made based on a final ranking that is statistically significant, taking into account criteria that are important to the user. For this reason, the procedure must be flexible in order to allow users to choose the criteria applicable to their particular requirements and deliver a practical outcome. The following results demonstrate these valuable characteristics. Furthermore, the results show what we think are generally valid rankings of these six DEMs.

\begin{figure}
\centering
\includegraphics[height=0.9\textheight]{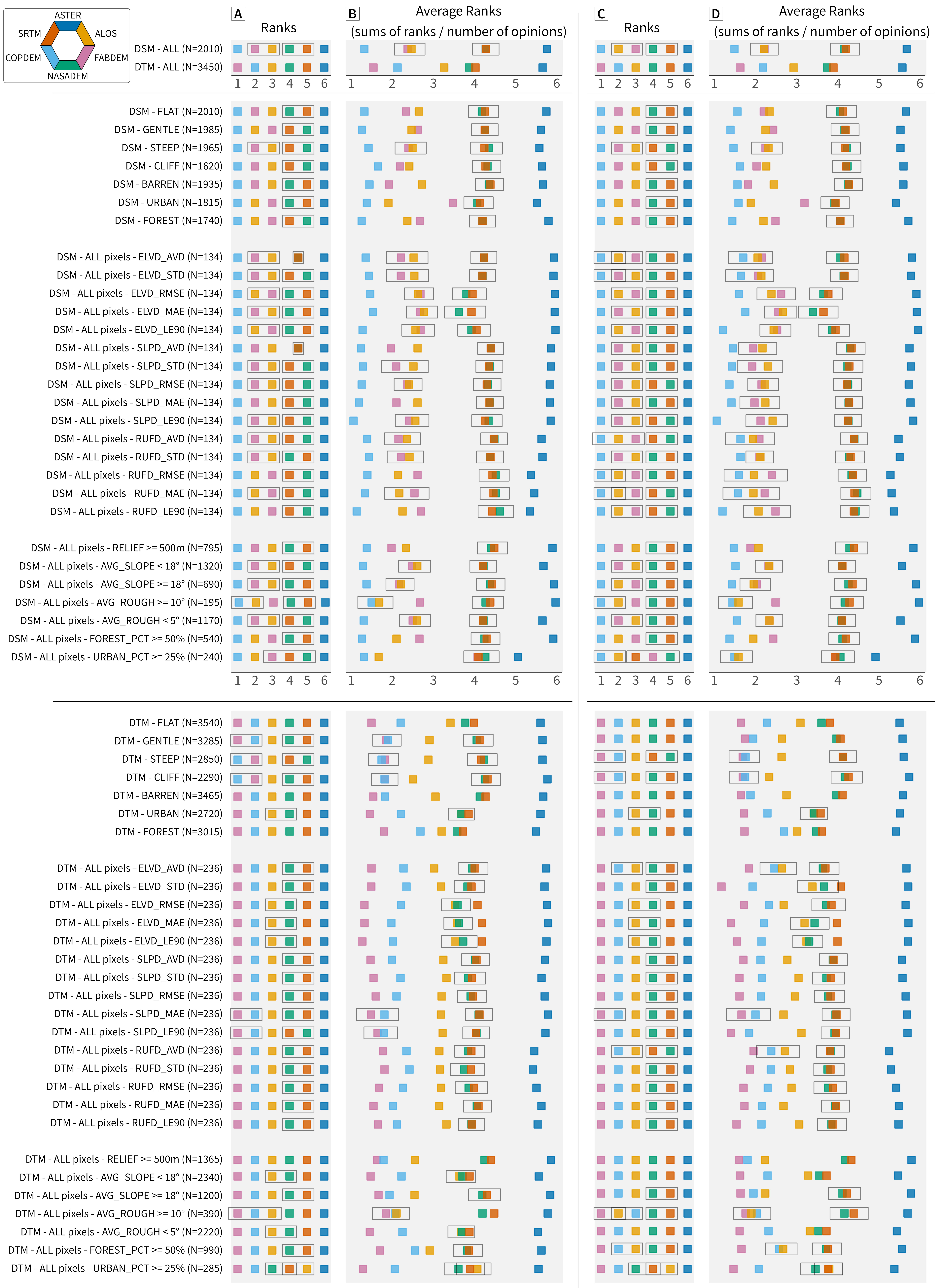}
\caption{The final rankings of DEMIX based on the inter-comparison of six 1'' global DEMs. The different outcomes are based on the chosen land types, the 15 contest criteria and filters for tiles with specific characteristics. Tolerances for A/B: ELVD=0.5, SLPD=0.5, RUFD=0.2. Tolerances for C/D: ELVD=1.0, SLPD=1.0, RUFD=0.4; the increase in tolerances led to a greater number of ties in C/D compared to the rankings shown in A/B.\label{fig:6_wc_full}}
\end{figure}

The output of the Jupyter notebook is illustrated in Figure~\ref{fig:6_wc_full}. The top of Figure~\ref{fig:6_wc_full} shows the overall ranking using all tiles and criteria based on the DTM and DSM reference DEMs, and the lower portion shows various filters of the database. The statistical significance to the ranking is presented in Figure~\ref{fig:6_wc_full} by drawing a box around those global DEMs whose rank cannot be differentiated from a random result, i.e., the pairwise rank outcome does not pass the significance test as such ranking outcomes are essentially ties. For the overall set of evaluations (ALL land type), the DTM winner is FABDEM and the DSM is CopDEM, both at the 95\% confidence level. For both the DSM and DTM ALOS is ranked only slightly lower, and for some comparisons such as DSM with high average roughness, ALOS is tied with CopDEM for first place.

In Figures~\ref{fig:6_wc_full}A/C, the inter-comparison rankings are ordered from best (1) to worst (6). From a statistical perspective, these rankings are ordinal and do not provide any information about how much better or worse any of the candidate DEMs are compared to each other. This is the expected output from DEMIX because it allows users to easily understand the best opinion based on their chosen criteria. A good impression of the variety of results is gained when analyzing the other contest rankings based on the subsequent rows. These have been grouped by reference DEM (DTM or DSM), land type filter (FLAT, GENTLE, FOREST, etc), and criterion (ELVD\_RMSE, SLPD\_LE90, RUFD\_MAE, etc.). Not only the order can change but also the number of ties and between which candidate DEMs the ties have occurred. This illustrates the power of the procedure in the context of the inter-comparison of global DEMs because depending on the user requirements, the `best' DEM will emerge based on the set of chosen criteria, far from a situation of one-option-fits-all.

To try and visualize the final ranking outcomes with a bit of context, Figure~\ref{fig:6_wc_full}B presents the same outcomes with the goal of showing how many times a test DEM ranks higher/lower over the number of opinions. The output is the sum of the ranks for any particular test DEM for each row in the opinions table (denoted as $R_j$ in Sec. 3), divided by the number of opinions. This visualization shows that the test DEM does not always have to be ranked number 1 in the opinions table to be the overall winner. For example, FABDEM is not the best DEM in very steep terrain. Columns A/B and C/D differ in showing the rankings on the left, and the scores on the right which provide a qualitative indication of how different the DEMs are. For instance ASTER is always the worst, but it is much worse than NASADEM and SRTM, which are actually very close. 

Figures~\ref{fig:6_wc_full}C and \ref{fig:6_wc_full}D demonstrate the effects that changing tolerances can have on the final rankings. The increasing of the tolerances produces more ties between the different global DEMs, but actually does not produce a tie for the first place DEM very often. This is an important result for the inter-comparison of global DEMs.

The results presented in Figure~\ref{fig:6_wc_full} are not exhaustive and help to demonstrate the power and flexibility of the RCBD method for the inter-comparison of global DEMs. The results also provide an important overview of the global DEM leaders based on the criteria used in this exercise: elevation differences, slope differences, and roughness differences.

\clearpage

\subsection{DEMIX Tile Results}
\label{sec:tile_results}

The final rankings presented in Figure~\ref{fig:6_wc_full} integrate almost forty thousand opinions in the DEMIX database (Table~\ref{tab:workflow}) which have been produced from 236 DEMIX tiles in 24 areas. These visualizations support the overall results produced through DEMIX and provide specific outcomes that support the results of the inter-comparison of the global DEMs.

\begin{figure}[!h]
\centering
\includegraphics[width=0.95\textwidth]{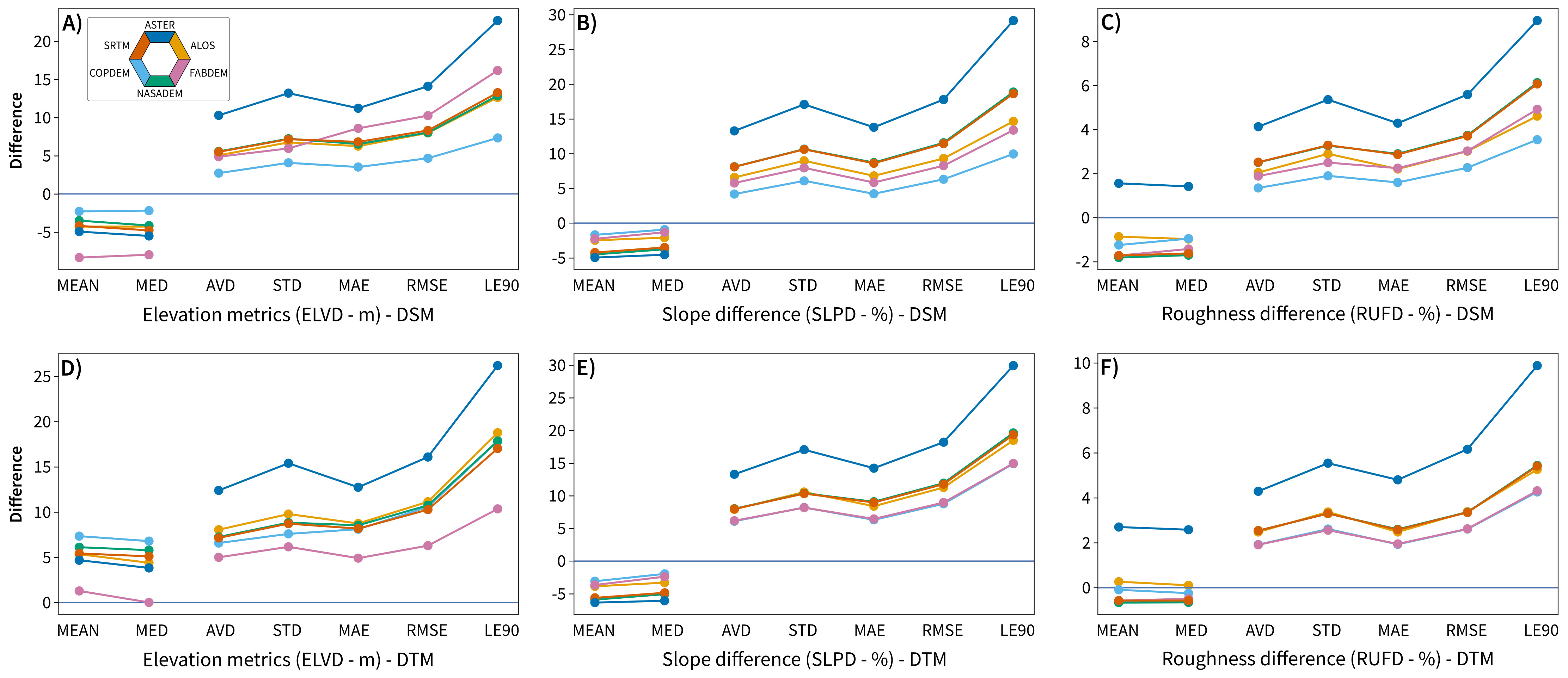}
\caption{The distribution of seven elevation, slope and roughness metrics comparing the difference between the test and reference DSMs (top) and DTMs (bottom) for tile N43PW002B in Spain. The mean and median on the left of each graph are signed; only the five rightmost parameters are used in the rankings. The lines connecting the points, while not statistically valid, show the effect on the parameters by their weighting of the tail of the distribution.\label{fig:7_metrics_spain}}
\end{figure}

Figure~\ref{fig:7_metrics_spain} displays seven metrics based on elevation (Fig.~\ref{fig:7_metrics_spain}A/D), slope (Fig.~\ref{fig:7_metrics_spain}B/E), and roughness (Fig.~\ref{fig:7_metrics_spain}C/F) differences between the reference and test DEMs for tile N43PW002B in Spain. The first two results displayed in each case (left side of each graph) are the mean and the median differences. These signed results cannot be ranked and therefore are not directly applicable; however, they provide valuable summary statistics for the tile under consideration. Specifically, they show how candidate DEMs compare to the reference DEM in terms of the direction of the differences: positive or negative. The remaining results are those used in the preparation of the evaluations table. Note that we show only one tile as a representative scenario of the metrics. While this is not a standard graphic, we show it for two important reasons.  First, there is little crossing of the lines, and the ranking of the DEMS is the same regardless of the criterion used--ASTER is always the most different from the reference DEM.  Second, popular metrics like MAE, RMSE, and LE90 produce different numerical results, but consistently lead to the same ranking of the DEMs.

For the signed results, one observes some negative bias for results based on the elevations (when compared to a reference DSM), slope, and roughness parameters. ASTER results tend to be outside of the group of results being an outlier in terms of not corresponding with the reference DEM compared to the other global DEMs, as clearly shown in Figure~\ref{fig:6_wc_full}.

The five unsigned metrics used as criteria show similar trends across the chosen tile and within each geomorphometric parameter. It would, however, be difficult to choose the best global candidate DEM based only on these graphs, especially with any statistical confidence.

\clearpage

\begin{figure}[!h]
\centering
\includegraphics[width=0.9\textwidth]{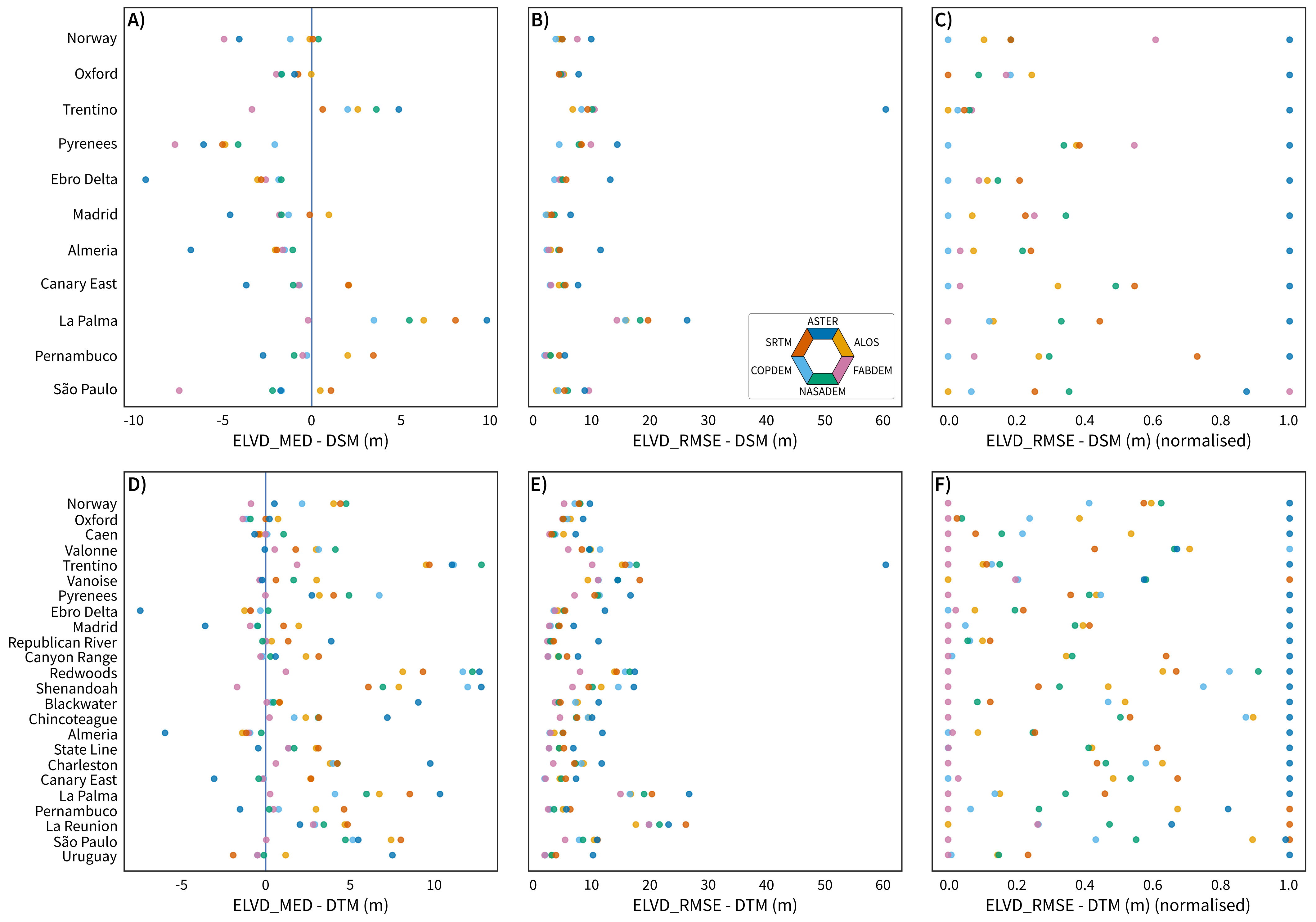}
\caption{Distribution of two signed metrics for all areas. A/D: ELVD\_MED. B/E: ELVD\_RMSE. C/F: ELVD\_RMSE with normalised values. Negative values for the median indicate the DEM is lower in elevation than the reference DEM, while positive values indicate the DEM has higher elevations than the reference DEM. The areas are arranged by latitude from north to south.\label{fig:8_signed_unsigned_all_tiles}}
\end{figure}

Figure~\ref{fig:8_signed_unsigned_all_tiles}A and~\ref{fig:8_signed_unsigned_all_tiles}D shows the medians of the elevation difference distributions for the 24 test areas, which average the results from 1--35 tiles in each area, highlighting a general negative bias for results based on the elevations when compared to a reference DSM and a positive bias compared to the DTM. Figure~\ref{fig:4_histograms_stateline} showed histograms with the distribution of elevation, slope, and roughness for a single tile, and Figure~\ref{fig:7_metrics_spain} showed all 7 parameters we computed for a different tile with elevation, slope, and roughness parameters. In all of these figures, and in virtually all comparisons we ran, ASTER results tend to be well outside the group of other results. 

Based on the DEMIX final rankings, it is apparent that CopDEM and FABDEM performed best in the global DEM inter-comparison, with ALOS a close third. However, such a conclusion could not be readily made based only on the visualization of the different metrics, especially as the number of locations increases. Unlike typical DEM comparison exercises, the DEMIX RCBD method provides statistical significance to the conclusions.

\section{Discussion}
\label{sec:discussion}

This paper has three objectives: (1) introduce the DEMIX RCBD method, (2) validate an open source tool chain to apply the methodology, and (3) apply the method for the first time to an inter-comparison of six global 1'' DEMs over a wide range of test areas with general recommendations on which of the DEMs performs best.

The results of DEMIX produced a ranking with prescribed confidence level of the global DEMs based on a set of quantitative criteria. The quantitative assessments were based on popular metrics applicable to intrinsic and derived parameter values of DEMs such as elevation, slope, and roughness.

The different contest final rankings produced a variety of results depending on the choice of criteria. For the DTM rankings (Fig.~\ref{fig:6_wc_full}), the leaders (top three) in order were FABDEM, CopDEM and ALOS. For some filtered criteria choices in the lower part of Figure~\ref{fig:6_wc_full} there was also a number of ties between these top three. NASADEM even moved into ties for third place in some particular cases where ALOS does not compare well with FABDEM and CopDEM. Overall NASADEM slightly improved SRTM, but only for limited land types and mostly for elevation metrics and not slope or roughness. For the DTM (lower tolerances) contests, ASTER, NASADEM and SRTM were always in the lower three places with ASTER always being clearly last.

The DSM rankings results differ from those of the DTM rankings both overall and across the different criteria. While the overall leader was CopDEM, ALOS and FABDEM share second and third place, sometimes tied. NASADEM and SRTM generally tied for fourth place; however, ASTER remained in sixth place throughout the DSM contests as well.

Varying tolerances (Figure~\ref{fig:6_wc_full}C/D) generally increased the number of ties but did not change the general trends observed in the DSM and DTM contest, and barely changed the first place results of CopDEM compared to a reference DSM and FABDEM compared to a reference DTM.

\subsection{Comparing the DEMIX Method and Rankings to Previous Work}

This work differs quite significantly from previous studies in a number of ways and most importantly in the total combination of factors that were applied for the inter-comparison and analysis of the candidate global DEMs. The following list compares and contrasts the DEMIX method to previous work:

\begin{enumerate}

\item Computed distributions of the differences between the reference and global DEMs for elevation, slope, and roughness. Many earlier studies just compare the distributions of the DEM with respect to the distribution of reference elevation data. Furthermore, the derived geomorphometric parameters in this study increase the effect of differences among the DEMs and the derived characteristics are often more important than the original DEM elevations for many applications.

\item Integrated seven parameters for each difference distribution -- most previous studies integrate fewer parameters typically only RMSE and MAE \cite{Grohmann2018}. While the chosen seven parameters are correlated, all the relationships are included in the database and show how little difference the choice makes.

\item Only 1'' DEMs were considered; changes in DEM spacing impact DEM characteristics making reliable inter-comparisons across multiple resolutions difficult. There are enough good quality global 1'' DEMs applicable to land-based processes that one should only consider 3'' DEMs such as MERIT \cite{Yamazaki2017} when lower spatial resolution is explicitly required or if the user feels that MERIT resembles a better DTM than FABDEM aggregated to 3''.

\item Created reference 1'' DEMs from the fine spatial resolution source reference DEMs, based on open data from national/local mapping agencies. These DEMs have significantly better spatial resolution and detail compared to the target 1'' resolution. However, the choice to use 1'' reference DEMs requires accepting this resolution and any limitations it imposes on the subsequent analysis.

\item No re-interpolation of the candidate global  DEMs was made after downloading other than adjusting to the EGM2008 vertical datum. Interpolation cannot improve a DEM or create additional elevations that were not captured in the original data.

\item All computations were performed in native geographic coordinates \cite{Guth2021}. While there are differences in the statistics between the rectangular 1'' pixels and square 30 m pixels, interpolating to UTM coordinates adds its own changes to computations done on the DEM which we avoided.

\item Created reference DEMs to consider the half pixel offset between ALOS and CopDEM (and the other lower quality DEMs) for the inter-comparison, a consequence of the decision not to resample candidate DEMs. This issue is generally not discussed in the literature (or addressed by interpolation) although \cite{Purinton2021} designed their comparisons to avoid the need for DEM co-registration.

\item Both DSM and DTM reference data are considered separately. It turns out that in most of our test areas, those not dominated by urban or forested land cover, the DTM and DSM vary little. Some of the minor differences arise from the different characteristics of the source data for the reference DEM: for example in Spain, these have different resolutions of 2 m and 5 m respectively, and were created at different times.

\item The inter-comparison was pixel-by-pixel at every point in the sample tiles, providing approximately 130,000 points per tile. This is a significantly greater number of points compared to studies using geodetic benchmarks, airport runways, or even measurements from laser altimeters such as GEDI and ICESat-2. In addition to providing often orders of magnitude more points for the inter-comparisons, the pixel-by-pixel option provides data in both steep and forested parts of the test area, where point-based reference data are often not available \cite{Gesch2014}.

\item A large number of diverse test areas -- in comparison, for example, \cite{Uuemaa2020} reported using 608 million 1 m pixels (after interpolation of the 1'' test DEMs), which corresponds to only six DEMIX tiles. \cite{Purinton2021} on the other hand used the equivalent of 400 contiguous DEMIX tiles, but from a single mountainous test area.

\item The DEMIX inter-comparison contest covered a wide range of land types (Fig.~\ref{fig:2_tiles_class}) which, combined with the large number of test areas, greatly increases the validity of the presented results. In contrast, some studies, such as \cite{Gesch2018} and \cite{Hawker2022}, focused their analysis primarily on coastal or floodplain terrain, which probably accounts for our finding that FABDEM often ranked lower than the CopDEM from which it was derived, particularly in mountainous terrain. At the other elevation extreme in mountain areas, \cite{Purinton2021} preferred CopDEM while \cite{Florinsky2019} and \cite{Trevisani2023} found that relative performances between CopDEM and ALOS were dependent on slope and roughness, with in general better performance of CopDEM except for very steep slopes. 
\end{enumerate}

Previous work did not quantify the ranking as developed within the DEMIX RCBD method. Our approach quantifies the inter-comparisons with additional metrics beyond RMSE or MAE and also considers elevation, slope, and roughness. Furthermore, the DEMIX methodology was applied to a representative land type sample covering a total  area of 23,600~km\textsuperscript{2} across three different continents. The closest approach to a DEMIX methodology applied to DEM characterisation is Table 4 in \cite{Dobre2021}, where they created a ``confusion matrix'' that ranked the DEMs from best to worst on four criteria.

\subsection{A DEM User Driven Contest}
A DEM user understands how to apply a DSM or DTM in their application, but is not necessarily familiar with how different DEMs were created or the different characteristics which can affect a user's DEM choice. The DEMIX methodology was developed to help DEM users make more informed decisions about the elevation data that they plan to use for their particular domain. This is a user-oriented method where experts can contribute with novel criteria or provide reference data to be shared with the community to improve user choices. The outcomes from this inter-comparison experiment demonstrate this ability of the method through the numerous rankings of the different criteria and regions of interest.

After DEMIX was started it soon became evident that equally challenging to testing and comparing DEMs at global scale would be to present the volume of results generated. Raw quality and validation figures alone usually require expert judgement on whether and to what extent a DEM is suitable for a certain application. The methodology offers a transparent and versatile solution to this problem. Not only does each criterion require that the results can be ranked, the real advantage to users comes with the possibility to combine criteria and test areas in any sensible way. This produces, in a transparent and statistically sound manner, at guiding recommendations on which DEM to use where and for what. For example, DEM experts could advise on which criteria to use and users could then rank available DEMs for the area or type of terrain they are most interested in.

Our study gives a first ordered ranking of DEMs. As further criteria and test areas are added to the database by the community, more comprehensive rankings and coverage of additional applications will be possible. However, we expect our general conclusions to remain valid.

How can DEM users take advantage of this wealth of scientific information? The DEMIX initiative is rectifying this issue with methodology that provides open access to high quality scientific results and provide DEM users:

\begin{itemize}
\item The ability to choose regions and locations based on standardized sampling areas (DEMIX tiles);
\item Allow users to choose relevant criteria to characterise and rank their DEMs of interest and;
\item The possibility to change the number of tiles as well as the criteria to produce new rankings among their DEMs of interest for comparison.
\end{itemize}

\subsubsection{Thematic Applications of the DEMIX Methodology}
The DEMIX methodology, including its flexibility and the possibility to evaluate DEM quality according to \textit{ad-hoc} chosen criteria, can be relevant in a multitude of application contexts, such as the study of geo-environmental processes, including the critical area of natural hazards. In the analysis and modelling of natural hazards, DEM derivatives are routinely adopted as geo-environmental proxies or as input features in supervised learning approaches \cite{Wilson2000,Hengl2008,Florinsky2016}. For example, a basic terrain variable such as slope represents a key factor in landslide susceptibility models, \cite{Titti2021,Huang2022}, as well as in the context of seismic hazard, where it has been adopted for the derivation of a proxy of shear wave velocity \cite{Wald2007,Heath2020}. Clearly, the possibility to evaluate and compare which DEM is better according to the representation of slope can provide insightful and useful suggestions for the users. For example, the results of the present DEMIX experiment suggest that CopDEM and ALOS should be preferred over the usually adopted SRTM and ASTER DEMs. The same discussion can be true in many other earth sciences related issues and topics such as fluvial morphology \cite{Boulton2018}, geodiversity studies \cite{Chrobak2021}, flood hazards \cite{Schumann2018}, and other domains.

No temporal criteria were taken into account during this first ranking of global DEMs, such as the date on which the measurements were taken or the length of time it took to acquire the global coverage. Temporal differences between global DEMs could be relevant to thematic applications especially if those changes occurred in the last 20 years and produced significant changes to the surface. Examples where significant elevation change could have occurred include volcanic eruptions, landslides, deforestation, urbanization, open pit mining, or severe erosion. The comparison should be carried out over comparable DEMs, and users must decide on what is comparable. If the user thinks that higher quality, newer DEMs are not representative of temporal changes under investigation, they can simply use the older DEMs. DEMIX proposes a paradigm shift, moving the focus from data producers to users. Traditional comparison exercises were expected to produce a definitive ranking, likely to be carved in stone. Now we expect to bring the power to the user, who will be able to justify their own choices with sound statistical arguments. 

DEM users should also be aware that reference data used in this work were not acquired at the same time as the global DEMs under consideration. The data collection date for the reference data used is provided in the metadata.

\subsubsection{Contest Criteria Ranking Requirements}
One of the powerful features of the methodology is the fact that users can define their own criteria. The only requirement is that the chosen metric must be designed and defined in such a way that the final outcomes can be ranked from best to worst/most appropriate to least appropriate. For example, in the above discussion it was mentioned that differences in elevation between the test and reference DEM, even if they contain useful information, cannot be used directly for ranking because these are signed quantities. 

For this reason, five criteria were based on the absolute differences of elevation instead. Strictly speaking, the adoption of the absolute value of differences implies a symmetrical weighting scheme in which the same importance is given to positive and negative elevation differences. However, in some applications, such as coastal flood risk analysis \cite{Kulp2018,Kulp2019}, negative and positive errors can impact our evaluations differently, for example in terms of cost functions. Accordingly, signed differences can be transformed  into unsigned and ordered rank quantities, weighting positive and negative errors differently (i.e., non symmetrically), taking advantage of a suitable transforming function.

A major rationale for developing the methodology, and indeed for informally calling it a wine contest, was the ability to include subjective criteria from expert judges. While this study did not include any subjective criteria, \cite{Guth2023_iasi_hillshade} compared hillshade maps of these 6 DEMs in 3 of the DEMIX tiles included here. In addition to validating our conclusions about the relative rankings of these DEMs, they point out the logistical testing challenges in setting up a subjective ranking for just 3 tiles, let alone the 236 tiles used for this work.

\begin{figure}
\centering
\includegraphics[width=0.55\textwidth]{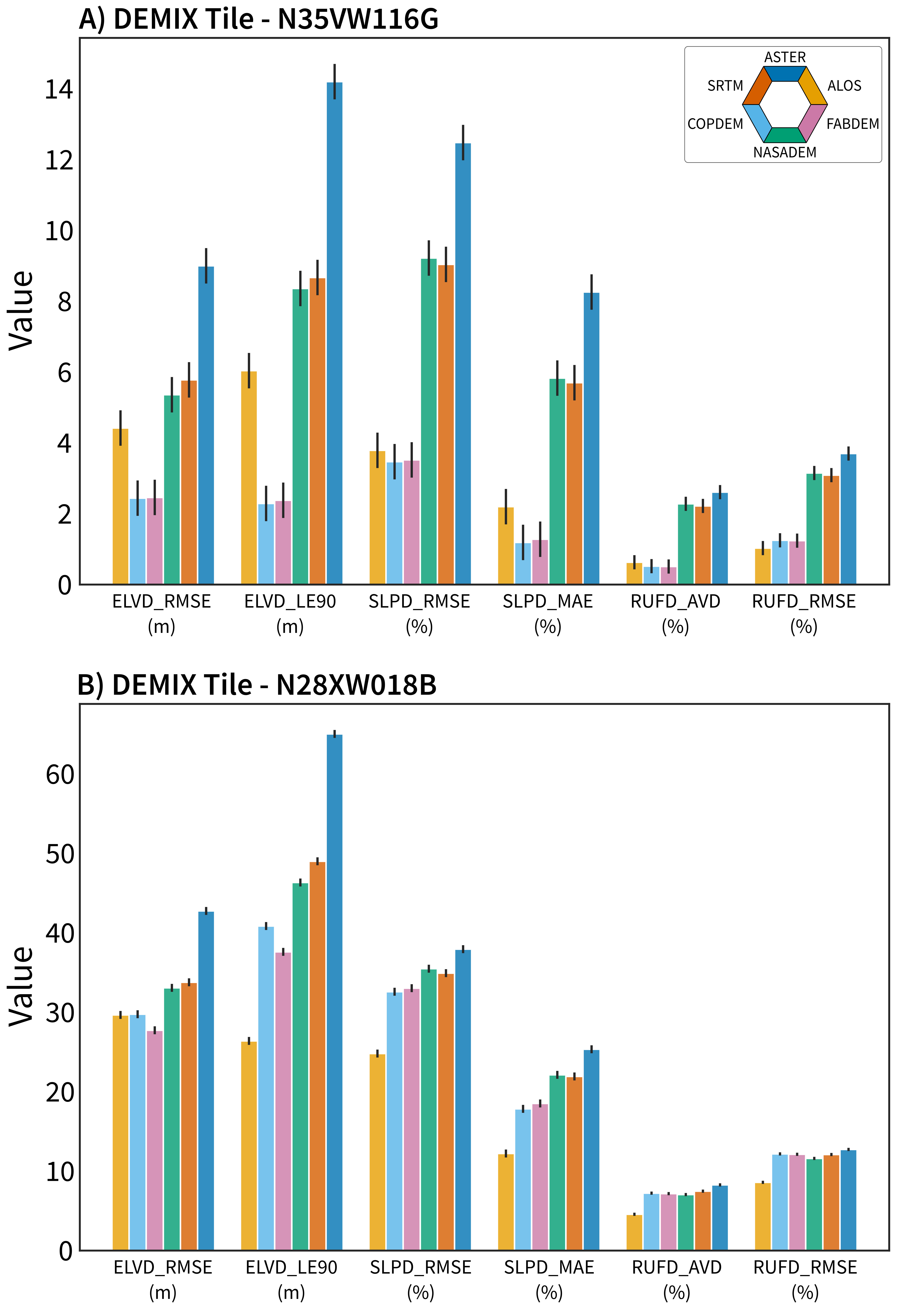}
\caption{Bar plots of the evaluation results for 6 criteria with two tiles compared to a reference DTM. The small black line atop each bar shows the tolerance we used for that criterion; and show the evaluation another candidate DEM would need to be tied.\label{fig:9_barplots}}
\end{figure}

Figures \ref{fig:7_metrics_spain} and \ref{fig:9_barplots} provide some guidance on the selection of criteria, and the tolerances for accepting times. In figure \ref{fig:7_metrics_spain}, the criteria used for the intercomparison on the right are separated from the signed means and medians. The five unsigned criteria are arranged from left to right in rough order by the effect of outliers; LE90 always has a much larger value compared to the others. With rare exceptions the lines connecting the values do not cross, and the rankings for the criteria are the same. The tolerances used for ties apply to the difference scale on the vertical axis; the amount of ``white'' space between the lines shows the tolerance that would be required to consider the DEMs tied. In most cases visual assessment of graphs like this reinforces our conclusions about the rank ordering of the DEM: CopDEM and FABDEM, then ALOS, then NASADEM and SRTM, and finally ASTER.

Figure \ref{fig:9_barplots} shows the evaluation results for 6 criteria and two different tiles. Note the very different vertical scales; tile N28XW018B in the Canaries is very steep and rough. Scores for the different criteria are also very different, although this also depends on the local topography. Salient results from this graph are that most of our results are immune to any reasonable choice of tolerances or which criteria are used, and that our ranking of the DEMs is robust. Tile N28XW018B is anomalous in showing ALOS as the best DEM for 5 of the 6 criteria; the results for tile N35VW166G are more typical (Table~\ref{tab:wine_contest_example}).

\subsection{Practical Application -- Hydrology}

Most users want to use DEMs for a wide range of practical applications, ranging from modeling landslide susceptibility, making flood models, to creating hillshade base maps. Statistical measures such as the 15 criteria used in our rankings do not always intuitively relate to what users want to apply, and most comparisons in the scientific literature use only a few specific metrics comparing elevations to reference data. We have argued that our combined assessment of elevation, slope, and roughness of the DEMs results in a more comprehensive and broadly applicable ranking of global DEMs compared to using only elevation metrics focusing on small regions.

To demonstrate the relevance of the wine contest results, we briefly look at a common application of DEMs in this section, and show that the global DEM rankings produced can also apply to the creation of a drainage channel network. For this demonstration, the test area contains 35 DEMIX tiles located near Madrid, Spain. Because channel networks cover larger areas than the 100 km$^2$ DEMIX tiles and channels can come into and out of individual tiles, using larger areas provides a better channel network by considering more of the full contributing drainage area. The SAGA (version 9.3.0) open source software for geospatial analysis \cite{Conrad2015,SAGA} was used with two modules: ``Sink Removal'' and ``Channel network'', accepting all default parameters, and using as input our 1 arc second reference DTMs derived from the Spanish national 2 m data \cite{Guth2023_demixrefdems} for both pixel-is-point and pixel-is-area geometries. The resulting vector channel network was rasterized to a 1 arc second grid in MICRODEM filtering to the channels with PLOT\_ORDER$>$1. 

\begin{figure}
\centering
\includegraphics[width=0.5\columnwidth]{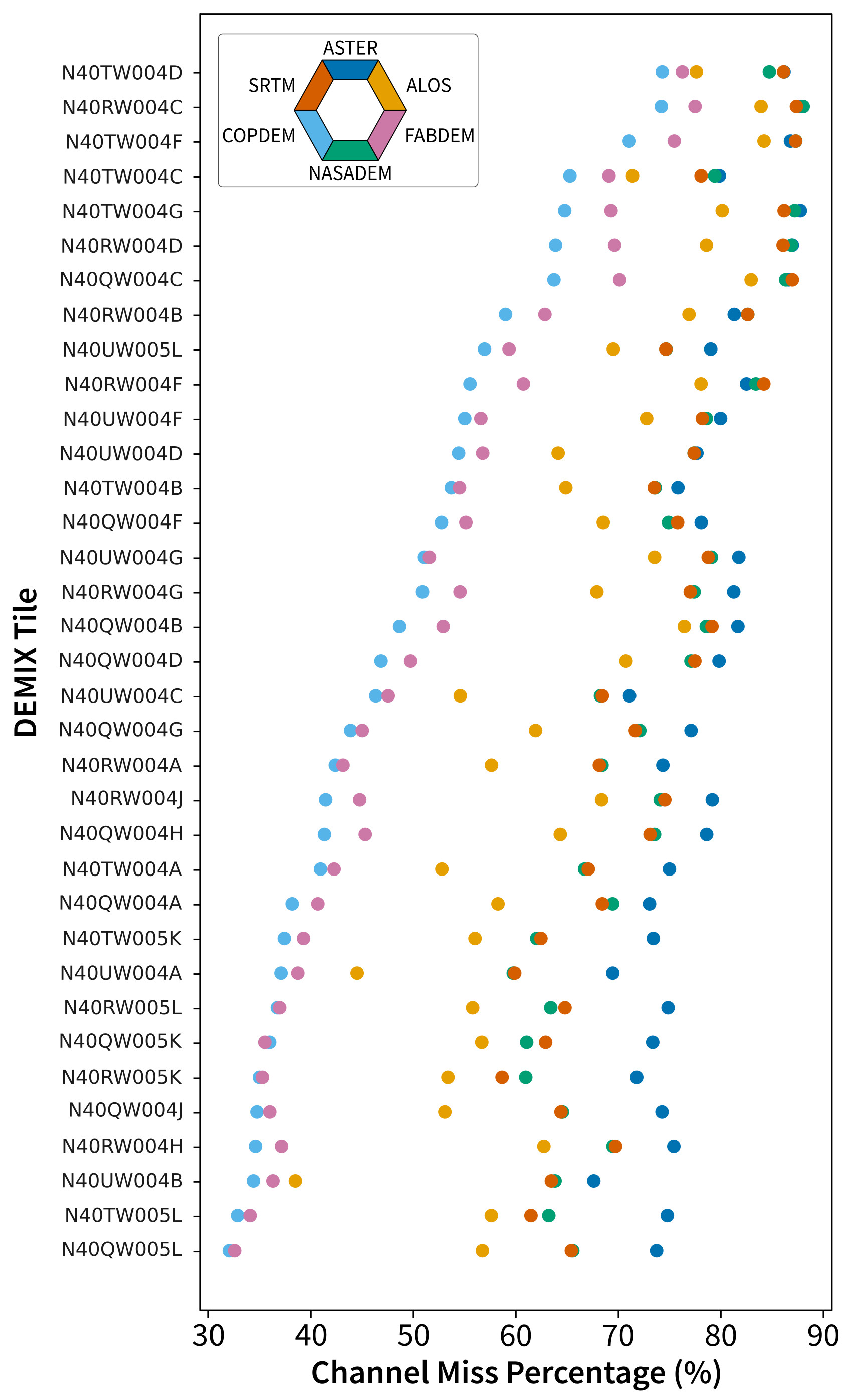}
\caption{Channel location comparison computed as the channel miss percentage for 35 DEMIX tiles in the Madrid (Spain) test area. A low score is better: 0\% means the channels derived from the candidate DEM perfectly match the reference DTM, and 100\% means they do not match at all.\label{fig:10_channels_ranks}}
\end{figure}

\begin{figure}
\centering
\includegraphics[width=0.95\textwidth]{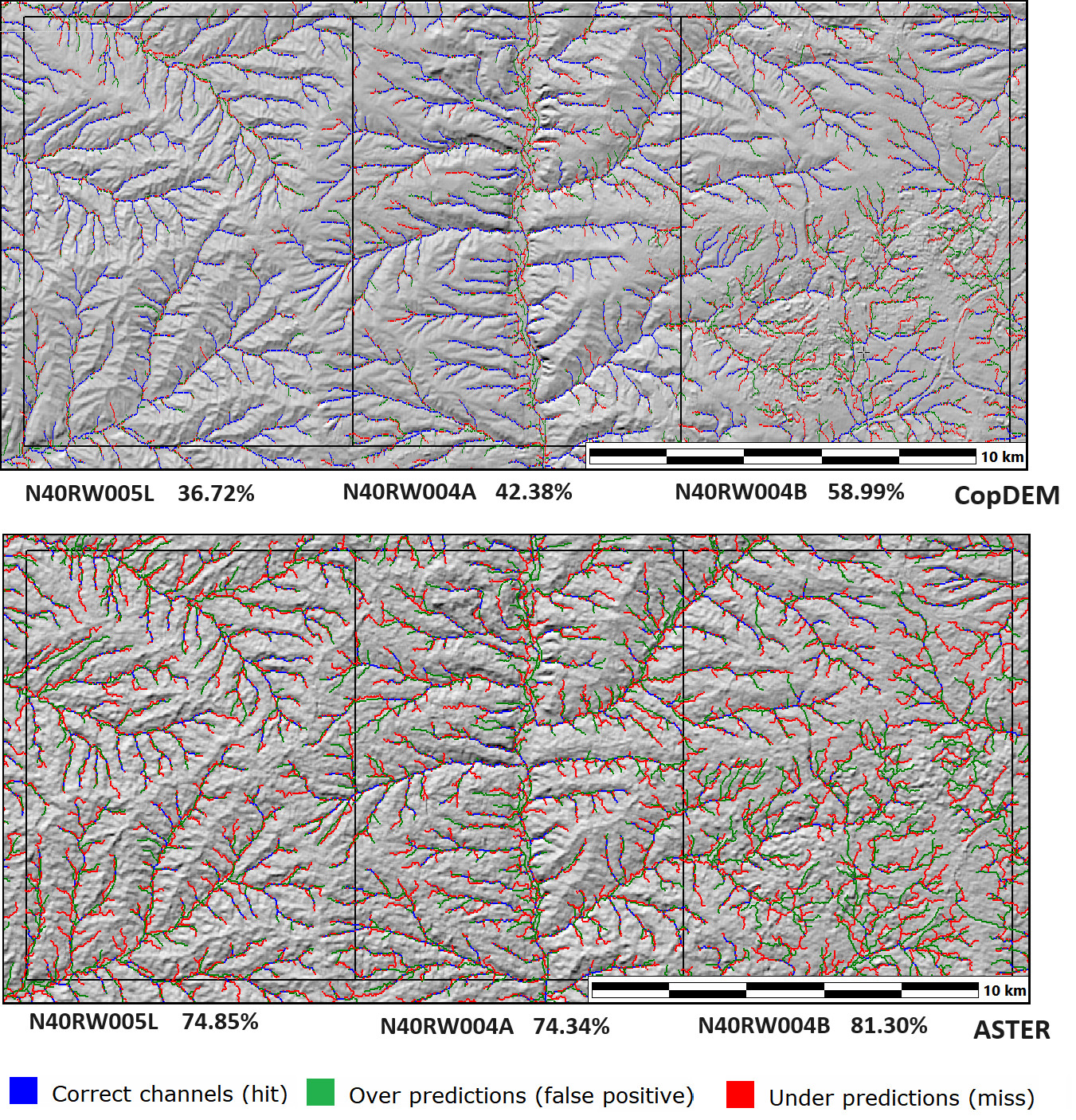}
\caption{Three tiles in the Madrid (Spain) test area, showing the miss percentages (low score best) for the channel  networks on hillshades for CopDEM (top) and ASTER (bottom). \label{fig:11_channels_shades}}
\end{figure}

Table~\ref{tab:channels} shows the four categories resulting from the analysis, and how the hit percentage (Correct channel / (Correct channel + Underprediction)) can be used to assess the results \cite{Stephens2013}. In figures \ref{fig:10_channels_ranks} and \ref{fig:11_channels_shades} we use a miss percentage (100 $-$ hit percentage) where lower is better which matches the criteria used in the wine contest and Figures \ref{fig:6_wc_full} and \ref{fig:8_signed_unsigned_all_tiles}.

\begin{table}[ht!]
    \centering
    \caption{Possible outcomes of the binary pattern measures computed between the pixels in the reference channel dataset and the six global DEMs.}
    \begin{tabular}{lll}
                     & Test DEM channel    & Test DEM inter-channel   \\
    \toprule
    Reference DEM    & Correct channel     & Underprediction  \\
    channel          &                     & (misses) \\
    Reference DEM    & Overpredictions     & Correct inter-channel \\
    inter-channel    & (false positive) \\
    \bottomrule  
    \end{tabular}
\label{tab:channels}
\end{table}

 Figure \ref{fig:10_channels_ranks} shows the results of comparing the river networks produced using the six global DEMs to the reference DTM and plotting the channel pixel miss percentage (low score is better). The relative order to the outcomes and the separations between the six global DEMs are constant for all 35 tiles. Differences among the tiles reflect the different terrain characteristics.

 A visual representation of the river network based comparison is shown for three tiles in Figure \ref{fig:11_channels_shades}. The top plot (CopDEM) shows the consistently lowest miss percentages (best depiction of the channel network), and the bottom plot (ASTER) presents an example of the DEM which consistently has the highest miss percentage (worst channel network). The easternmost tile, N40RW004B, is largely urban, and has the largest miss percentage compared to the reference DTM. Flat areas are a known problem in channel extraction \cite{Zhang2017}, and all DEMs are not good as seen in the urban tile on the right and the main river valley in the center tile. Even in the hilly areas, the ASTER DEM tiles that are shown miss more of the small order streams. In addition to the channel networks, the hillshade from ASTER is much rougher and does not show the ridges and valleys as well as that from CopDEM.

 The global DEM intercomparison that produced the rankings in this paper did not directly consider hydrology. The example presented in this section based on channel networks in the Madrid area closely matches the rankings of the six global DEMs in our holistic set of elevation, slope and roughness criteria. A qualitative assessment of hillshade maps for several DEMIX tiles had similar results and ranking of the six global DEMs \cite{Guth2023_iasi_hillshade}. Both the channel network and hillshade results are based on significantly smaller sample sizes, but we find the outcomes support the wine contest DEM intercomparison rankings (see figure \ref{fig:6_wc_full}) which we expect to be generally valid for a broad range of applications. This hydrology experiment verifies the applicability of the global wine contest rankings beyond having better RMSE or LE90 statistical values and the ranking suggests that most practical applications will benefit from using the DEMs recommended by our rankings.

\section{Conclusions}
\label{sec:conclusions}

This paper presents the novel `DEMIX wine contest methodology', for the inter-comparison of digital elevation models which produces a final ranking with prescribed confidence levels based on given criteria. We applied the method to six global 1'' DEMs: ALOS, ASTER, CopDEM, FABDEM, NASADEM, and SRTM. The inter-comparison was done using 15 criteria related to elevation, slope, and roughness measures derived from reference 1'' DEMs. The final rankings with confidence level based on the choice of criteria and land type, demonstrate the powerful features of this method, including the ability for the user to choose the most relevant criteria and areas.

The main outputs from the DEMIX global DEM inter-comparison are presented in Figure~\ref{fig:6_wc_full} and further supported by other figures included in the text to provide the experts and users transparency with respect to the following conclusions:

\begin{enumerate}
\item From an overall final ranking of the global 1'' DEM inter-comparison, CopDEM and FABDEM are clearly the frontrunners based on the chosen criteria and test areas: CopDEM when compared to a DSM and FABDEM when compared to a DTM. The best rankings depend on the choice of parameters and what the user wants to do with the DEM. ALOS was generally the third best, but sometimes moving into second place.

\item FABDEM improves on CopDEM in most cases where a DTM is required, except for steep terrain which was not an emphasis in its generation.

\item SRTM and NASADEM are distinctly in the lower half of the wine contest rankings indicating lower quality than the top three. Consequently, unless compelling reasons (such as surface changes) are presented, we encourage users to move away from using them. While SRTM revolutionized global digital topography, NASADEM only produced modest improvements and actually decreased scores for some criteria. The improvements in NASADEM are mostly for the elevation measures, and it did little to change the derived slope and roughness measures. 

\item As many prior studies have also shown, the ASTER DEM is clearly the lowest performer.

\item Certain areas show quite anomalous DEM rankings. For example, ALOS on average was the best in about 8.4\% of the opinions, but in four areas (Vanoise, São Paulo, La Reunion, and La Palma) it was better at least twice as often. This supports the DEMIX goal to help DEM users choose appropriate products for their areas of interest and the criteria most relevant to their usage.

\item All the global 1'' DEMs that were evaluated by DEMIX are composite, i.e., they required other data sources for in-filling gaps because the original data acquisition did not provide full-coverage. For example, CopDEM makes extensive use of high quality national lidar data for infilling and improving TanDEM-X data over Norway. Results in this area are therefore not representative for most other test areas.

\item The three high quality DEMs (CopDEM, FABDEM, ALOS) are all global and there is no longer an issue with the lack of  coverage at high latitudes. However, local availability of the main instrument data varies sigificantly among ALOS, CopDEM and FABDEM and may affect their spatial consistency. 

\item The three global high quality DEMs are much closer in time of collection: CopDEM and FABDEM: 2010-2015, and ALOS: 2006-2011. Except for SRTM, all data used to produce the global DEMs were collected over a number of years. The reference DEM data used here also come from a wide range of dates. Global DEM users requiring data from 2000 can take advantage of these results to decide whether SRTM or NASADEM meet their elevation needs.

\end{enumerate}

The breakout of the six global 1'' DEMs into two groups corresponds to the technologies and sensor characteristics used to acquire the elevation data. SRTM, NASADEM, and ASTER are delivered having a pixel size of 1'' but from the beginning it was recognized that the actual resolution did not match comparable cartographic DEMs and that the `real' resolution was coarser than 1'' \cite{Smith2003,Guth2006,Pierce2006,Reuter2009a}. In contrast, ALOS and CopDEM are derived from commercial DEM products having resolutions of 0.15'' ($\sim5$~m spatial resolution) and 0.4'' ($\sim12$~m) respectively. Consequently, ALOS and CopDEM have a resolution appropriate to their pixel size and therefore better match the reference DEMs.

In this experiment, CopDEM consistently ranks best when compared to the reference DSM, and FABDEM ranks best when compared to a DTM. \cite{Guth2021b} showed that CopDEM consistently had elevations tightly clustered near the center of the lidar point cloud, and thus was intermediate between a DSM and a DTM. The other DEMs investigated in that work had much greater dispersion in the point cloud. Histograms from the difference distributions (Figure~\ref{fig:4_histograms_stateline}) for all the tiles we examined support the general applicability of this generalization, and that it applies to slope and roughness as well as elevation.

\section{Acknowledgments}

Support to the DEMIX project is performed by the European Space Agency (ESA) in the framework of the EDAP+ project. CHG is supported by CNPq (grant \#311209/2021-1) and FAPESP (grant \#2019/26568-0). LH is funded by the Natural Environment Research Council (NERC) EVOFLOOD Project (NE/S015795/1). Any use of trade, firm, or product names is for descriptive purposes only and does not imply endorsement by the U.S. Government.

\appendix

\section{Global DEMs charateristics}
\label{apx:gdems}
\setcounter{table}{0}

\begin{table*}[ht!]
    \centering
    \caption{Free quasi-global DEMs at 1'' \citep[atfer ][]{Guth2021a}.}
    \begin{adjustbox}{width=0.99\columnwidth}
    \begin{tabular}{%
  >{\raggedright}p{3.6cm}
  >{\raggedright}p{3.2cm}
  >{\raggedright}p{2.6cm}
  >{\raggedright}p{2.8cm}
  >{\raggedright}p{3.0cm}
  >{\raggedright}p{3.0cm}
  >{\raggedright}p{2.8cm}
  p{2.5cm}<{\raggedright}}
    \toprule
    \textbf{DEM}                & \textbf{Primary Source}        & \textbf{Producer}  & \textbf{Vertical Datum}  & \textbf{Precision}         & \textbf{Grid Storage}  & \textbf{Longitudinal Spacing} & \textbf{Acquired}\\
    \midrule
    SRTM v3                     & C band radar                   & NASA               & Orthometric EGM96        & Integer                    & Pixel-is-Point     & Constant                      & 2000 (11 days) \\ 
    ASTER GDEM v3               & Stereo NIR imagery             & NASA/METI          & Orthometric EGM96        & Integer                    & Pixel-is-Point     & Constant                      & 2000-2013      \\ 
    ALOS World 3D AW3D30 v3.2   & Stereo pan imagery             & JAXA               & Orthometric EGM96        & Integer                    & Pixel-is-Area      & Variable                      & 2006-2011      \\ 
    NASADEM                     & Reprocessed C band radar       & NASA               & Orthometric EGM96        & Integer or floating point  & Pixel-is-Point     & Constant                      & 2000 (11 days) \\ 
    Copernicus DEM GLO30        & X band radar, Edited WorldDEM  & ESA/Airbus         & Orthometric EGM2008      & Floating point             & Pixel-is-Point     & Variable                      & 2011-2015      \\ 
    FABDEM v1-2                   & Edited Copernicus DEM          &                    & Ellipsoidal WGS84        & Floating point             & Pixel-is-Point     & Constant                      & 2011-2015      \\ 
    \bottomrule

    \end{tabular}
    \end{adjustbox}
\end{table*}

\subsection*{Global DEMs data sources}

\begin{itemize}
\item SRTM -- \url{https://earthexplorer.usgs.gov}, \url{https://search.earthdata.nasa.gov/search/granules?p=C1000000240-LPDAAC_ECS}

\item ASTER GDEM -- \url{https://ssl.jspacesystems.or.jp/ersdac/GDEM/E/}, \url{https://search.earthdata.nasa.gov/search/granules?p=C1711961296-LPCLOUD}

\item ALOS AW3D30 --\url{https://www.eorc.jaxa.jp/ALOS/en/aw3d30/index.htm}

\item NASADEM -- \url{https://doi.org/10.5067/MEaSUREs/NASADEM/NASADEM_SHHP.001}, \url{https://search.earthdata.nasa.gov/search?q=C1546314043-LPDAAC_ECS}

\item Copernicus DEM -- \url{https://doi.org/10.5270/ESA-c5d3d65}

\item FABDEM -- \url{https://doi.org/10.5523/bris.s5hqmjcdj8yo2ibzi9b4ew3sn}

\end{itemize}

\clearpage
\subsection*{Supplementary Material}

\begin{figure}[!h]
\centering
\includegraphics[width=0.99\textwidth]{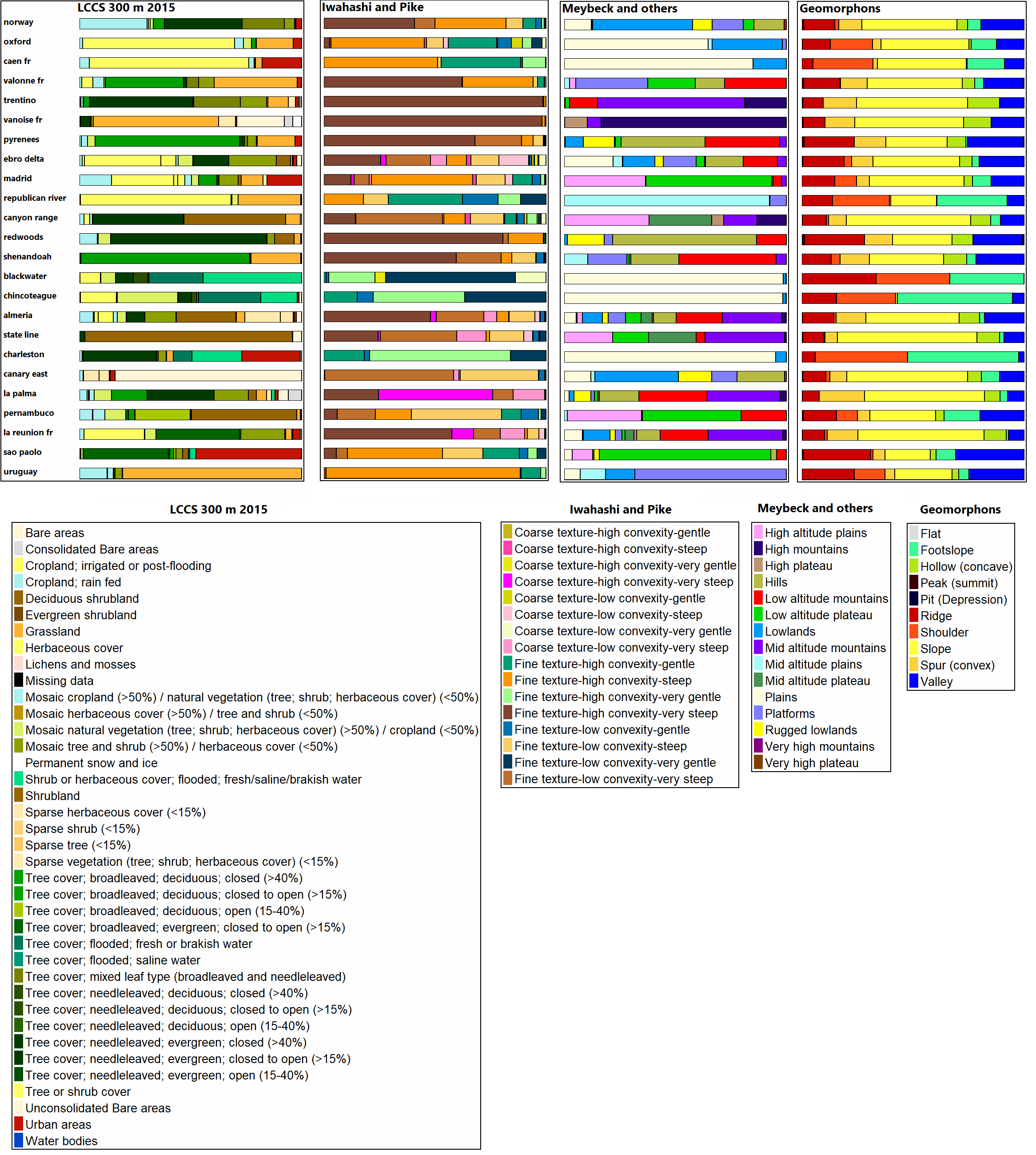}
\caption{Tile classifications for the 24 DEMIX test areas in our sample, for one landcover classification and 3 geomorphometric landform classifications. Areas are arranged from north to south. This is an expanded version of Figure 3 of the text to include legends.}
\end{figure}

\clearpage

\begin{figure}[!h]
\centering
\includegraphics[height=0.95\textheight]{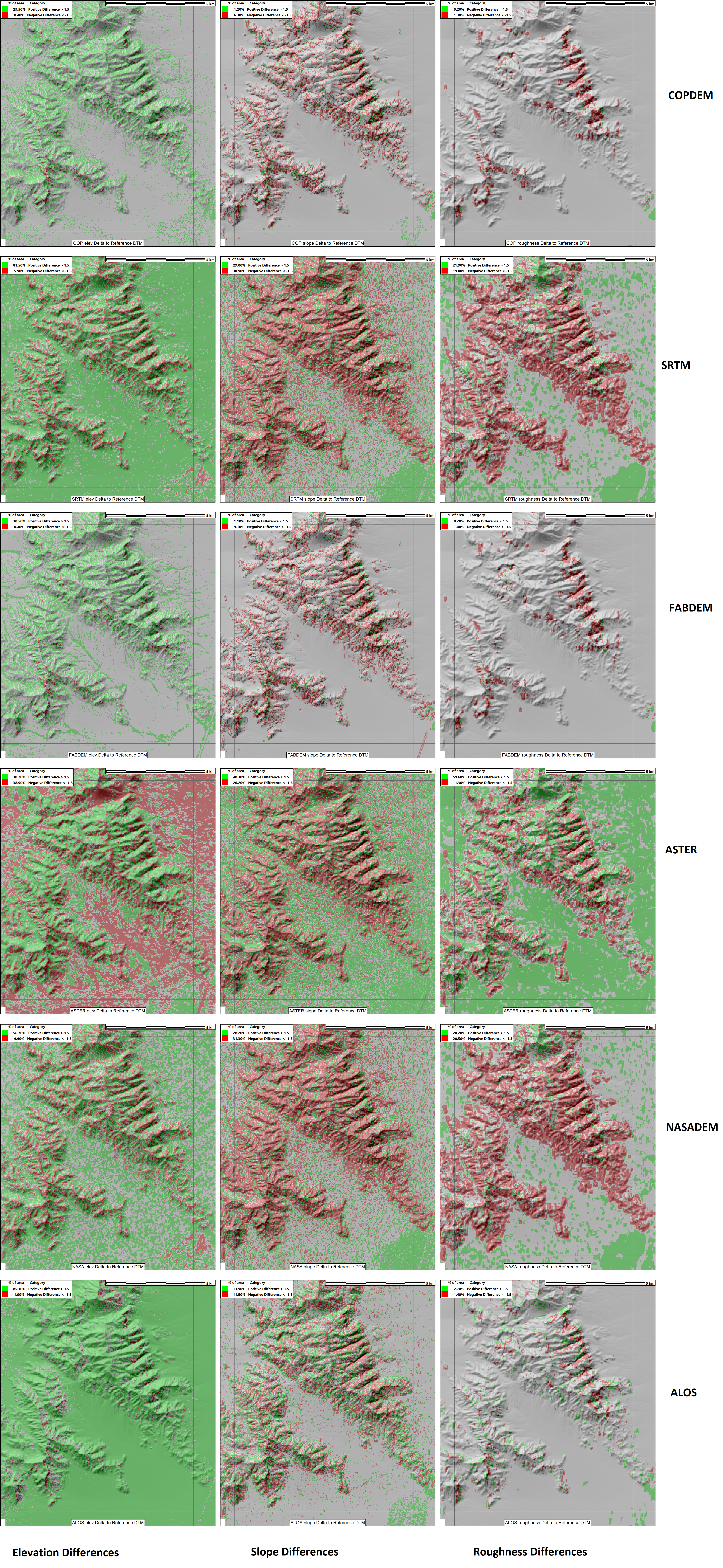}
\caption{Difference maps for the 6 global 1'' DEMs, by row, for the elevation, slope, and roughness by column.  The legends show the percentage of in green that are higher, steeper, and rougher than the reference DTM, and in red that are lower, less steep, and smoother.  This is an expanded version of Figure 4 of the text to show difference maps for all 6 DEMs.}
\end{figure}  

\clearpage


\end{document}